\pdfoutput=1
\documentclass[fleqn,10pt]{wlscirep}
\usepackage[T1]{fontenc}
\usepackage{subcaption}
\usepackage{makecell}
\usepackage[version=4]{mhchem}
\usepackage{siunitx}
\usepackage[capitalise]{cleveref}
\usepackage{placeins}

\newcommand{\SEDS}{\ensuremath{\Sigma\mathrm{EDS}}}
\newcommand{\qinv}{\ensuremath{\text{\AA}^{-1}}}

\title{Nucleation at the GST--oxide interface and beam-driven ordering in amorphous Ge--Sb--Te: HAADF, EDS and electron pair distribution functions across a 4D-STEM dose series}

\author[1,*]{Mridul Kumar}
\author[1]{Kirill Sobolev}
\author[2]{Alexander Upcher}
\author[1,2,*]{Yevgeny Rakita}
\affil[1]{Department of Materials Engineering, Ben-Gurion University of the Negev, Be'er Sheva, Israel}
\affil[2]{Ilse Katz Institute for Nanoscale Science and Technology, Ben-Gurion University of the Negev, Be'er Sheva, Israel}
\affil[*]{Corresponding authors: mridul@post.bgu.ac.il, rakita@bgu.ac.il}

\begin{abstract}
Where and why crystal nuclei form in amorphous phase-change materials decides how they switch. We study six
successive four-dimensional scanning transmission electron microscopy (4D-STEM) scans, with simultaneous
high-angle annular dark-field (HAADF) and energy-dispersive X-ray (EDS) data, of a magnetron-sputtered Ge--Sb--Te (GST)
film, restricted to the GST layer and its interface with the oxide layer beneath it. We first test whether nucleation sites differ
in composition or number density: they do not. The nuclei lie on the same robust HAADF--EDS relation as the matrix,
and at fixed HAADF the spread of the summed EDS signal equals its Poisson counting noise (ratio
\num{1.03 \pm 0.05}). The nuclei, marked by hand with the help of an automatic Bragg-spot detector, are however
crystalline: every one exceeds the detector's Bragg-contrast threshold, and they stand out in the 4D-STEM max-intensity virtual image
(median $z = 11.0$) but not in HAADF ($z = 1.3$) or EDS ($z = 0.3$). We then classify the film in four different ways, by HAADF intensity, by $I(q)$ phase mapping, by
principal-component analysis (PCA) of the whole diffraction patterns and by a self-supervised (SimCLR) foundational
encoder, and compare the electron pair distribution functions (PDFs) of all classes
after calibrating reciprocal space on the Au cap. All bulk classes share one short-range structure (first shell
\SIrange{2.70}{2.74}{\angstrom}); what changes is depth and dose. Towards the interface the first sharp diffraction
peak (FSDP) moves to lower $q$, and every method isolates a thin interface-edge environment with a shorter or
missing first shell. The $I(q)$ classification isolates, in every scan, an interface environment that holds
\SIrange{62}{100}{\percent} of the nucleus pixels (\numrange{1.2}{2.5}-fold above depth) and that persists when the
nucleus pixels are removed before classification; whole-pattern PCA does so in every scan and the HAADF bands in four of
six, while the self-supervised encoder separates the crystalline nucleus pixels from their surroundings. The
interface, rather than a distinct composition or density of the nucleating volume, therefore marks where nuclei form. Over
\SI{4.7e3}{e^- \per\angstrom\squared} the beam removes \SI{17}{\percent} of the HAADF signal, shortens the first
shell by \SI{0.03}{\angstrom} and narrows the FSDP by \SI{9}{\percent}, apparently ordering the film as it damages it. All
analyses used PySTEMLab, an interactive 4D-STEM analysis environment developed for this work, either interactively or
through scripts that call its core routines.
\end{abstract}

\begin{document}
\flushbottom
\maketitle
\thispagestyle{empty}

% ════════════════════════════════════════════════════════════════════════
\section{Introduction}

Phase-change materials (PCMs) based on Ge--Sb--Te (GST) alloys underpin non-volatile memory, neuromorphic and
optical storage technologies\cite{wuttig2007phase,raoux2010phase,jones2024pcm,lan2020thermophysical}. They store
information in the contrast between an amorphous and a crystalline state, and the speed, endurance and
retention of a device are set by how the amorphous phase crystallises: where nuclei form, how many form and how
fast they grow\cite{kelton2010nucleation,kalb2007calorimetric}. Interfaces, surface oxides and capping layers can change the
preferred nucleation sites\cite{agati2020surface}. Observing
nuclei at the stage at which they are only a few nanometres across, and asking what distinguishes the regions in
which they form, is therefore central to understanding and engineering PCMs.

Conventional characterisation struggles with this question. X-ray diffraction and conventional transmission
electron microscopy average over large volumes or give limited structural information about amorphous
regions\cite{epp2016xrd,ortega2022characterization,tang2017tem}. Energy-dispersive X-ray spectroscopy (EDS)
maps composition, but needs high doses and long acquisitions that can damage sensitive
samples\cite{hodoroaba2020eds,schmied2002particle}, and its single-pixel counts are small. High-angle annular
dark-field (HAADF) imaging is sensitive to the mean atomic number $Z$ and thickness. Four-dimensional scanning
transmission electron microscopy (4D-STEM) records a full diffraction pattern at every probe
position\cite{ophus2019fourdimensional,savitzky2021py4dstem} and so measures local order directly, but the
resulting high-dimensional data require dedicated computational tools\cite{sadri2024unsupervised,allen2021fast,
shi2022uncovering}.

In two earlier studies of the same GST film we predicted EDS maps from diffraction patterns with a convolutional
neural network\cite{kumar2025prediction} and clustered the diffraction patterns into chemically interpreted
classes\cite{kumar2025clustering}. Both left the nucleation question open: a crystalline region was read as a
prediction error, the clusters were interpreted through small differences in EDS intensity, and the interface
region, where the nuclei turn out to be, was cropped away.

Our starting hypothesis was that nucleation centres differ in atomic number density or composition from their
surroundings. Simultaneous HAADF and summed EDS (\SEDS) are the natural probes of that hypothesis: HAADF scales with
$Z$ and thickness, \SEDS{} with the number of atoms in the probed column, so nuclei of different density or
composition should fall off the HAADF--\SEDS{} relation of the rest of the film. The second probe is local atomic
structure, for which the pair distribution function (PDF) is the most direct measure: the reduced PDF $G(r)$
obtained from the diffuse scattering of amorphous material gives interatomic distances in real space, and the first
sharp diffraction peak (FSDP) of the scattered intensity reports medium-range order\cite{elliott1991mro,
kohara2006gst}. Electron diffraction gives PDFs from nanometre-sized volumes\cite{cockayne2007nanovolumes}, so
4D-STEM can provide a PDF for every class of probe positions, however the classes are defined.

Here we restrict the analysis to the GST layer and its interface with the oxide layer beneath it, classify the probe positions
by HAADF intensity, by $I(q)$ phase mapping, by PCA of the whole diffraction patterns and by a self-supervised foundational
encoder, and compare the PDFs of all classes. To make this possible we developed PySTEMLab,
an interactive environment in which every step, from region selection to PDF extraction, can be carried out,
inspected and reproduced. The Materials and Methods, including a short description of PySTEMLab, are followed by the
Results; secondary analyses are in the Supplementary Information (SI).

% ════════════════════════════════════════════════════════════════════════
\section{Materials and Methods}

\paragraph{Sample Preparation.} The \ce{Ge2Sb2Te5} thin film was deposited in a custom magnetron sputtering chamber
from a single \ce{Ge2Sb2Te5} target (MSE Supplies, USA), using an RFX-600 RF power source (Advanced Energy Industries
Inc., UK) operated at \SI{40}{\watt} for \SI{500}{\second} to reach the target film thickness of \SI{100}{\nano\metre}.
The deposition rate had been calibrated beforehand by depositing films at different powers and times and measuring their
thicknesses by atomic force microscopy (AFM). The substrates were lab-grade glass slides, cleaned by successive
sonication in acetone, isopropanol, Alconox\textsuperscript{\textregistered} detergent and double-distilled water
(\SI{15}{\minute} each at \SI{80}{\kilo\hertz}) and dried with a compressed-air gun. Before deposition, the target was
held under Ar plasma at the deposition conditions for \SI{10}{\minute} with the shutter closed, to remove possible
surface contaminants and reach a stable deposition rate.

\paragraph{Acquisition.} The same film (with an Au cap and an oxide-rich layer) was studied
previously\cite{kumar2025prediction,kumar2025clustering}; in the scans the Au cap lies directly on the GST and the
oxide layer is on the opposite side. Six successive scans (0033--0038, in acquisition order) of the same $110 \times 45$ pixel region (pixel size
\SI{1.053}{\nano\metre}, magnification \num{185000}, probe size \SI{2.5}{\nano\metre}) were acquired on a Thermo
Fisher Spectra 200 at \SI{200}{\kilo\volt} with a convergence semi-angle of \SI{0.46}{\milli\radian}, a dwell time of
\SI{100}{\milli\second} (frame time \SI{506}{\second}) and a recorded probe current of \SI{0.14}{\pico\ampere}. Each
probe position yields a $256 \times 256$ diffraction pattern, a HAADF value (\SIrange{138}{200}{\milli\radian}) and
an EDS spectrum from four Super-X detectors. Diffraction patterns were centred and the bright-field disc masked with
PySTEMLab's Shift \& Mask dashboard.

\paragraph{Analysis Window.} Rows 0--19 contain the Au cap (Au $\ge$ \SI{10}{at.\percent} down to row 8--11, depending on the scan, with
intermixing below) and rows 104--109 the beam-damaged oxide layer. Rows 20--103 (GST and the GST--oxide interface,
including every nucleus) were saved as a rectangular region of interest in PySTEMLab and used by every dashboard and
script, so that all classes and statistics refer to the same 3780 probe positions per scan.

\paragraph{HAADF--EDS Analysis.} \SEDS{} is the per-pixel sum of all counts above channel 74. The HAADF--\SEDS{}
relation is fitted by Theil--Sen regression with nuclei excluded; nucleus residuals are compared with those of matrix
pixels in the same rows by a two-sided Mann--Whitney test\cite{mann1947test}. The HAADF bands are a
stratification of the continuous HAADF axis into six slices of nearly equal HAADF intensity, used for the tests at
fixed HAADF; they are not classes found in the data. They were set with the band tool of PySTEMLab's correlation
explorer rather than drawn freehand: six bands of equal width in HAADF intensity, separated by gaps, starting at the
tool's default upper edge, the 99.9th percentile of the window's HAADF distribution. The band width and gap were set by
hand to round values that make the six bands span the HAADF range of the GST layer: 400 HAADF units with gaps of 80 in
scans 0033--0036, and 300 with gaps of 60 in 0037--0038, where beam damage has compressed the HAADF range
(Supplementary Section~S6 compares them with the tool's default bands). Pixels in
the gaps or below the lowest band (\SIrange{17}{20}{\percent} of the window) stay unassigned, so that neighbouring
bands do not share edge pixels. HAADF intensity scales with the number of atoms in the beam path, each weighted by
roughly $Z^{n}$ ($n \approx 1.6$--2), so a band groups probe positions of similar mass--thickness, not of similar
composition. Mixture models use standardised (\SEDS, HAADF), full covariance, 1--8 components, BIC
selection and five initialisations.

\paragraph{Bragg-spot Detection and Visibility.} Each pattern is smoothed ($\sigma = 1$ px); local maxima in a
$5 \times 5$ window at $16 \le r \le 120$ px define candidates, with relative contrast to the azimuthal mean at their
radius. Seeds exceed the scan median by six, and grown regions by three, robust standard deviations
($1.4826 \times$ MAD); objects of fewer than two 8-connected pixels are dropped. Visibility compares the contrast of
each object (object mean over a 4-px ring) with every valid placement of the same shape at similar depth (row window
$\pm 6$, widened in steps of 3 up to $\pm 24$ until at least 30 placements), at least 6 px from any manual or automatic object; objects, rings
and placements are restricted to the window. \emph{Nucleus masks:} the nuclei were marked manually in the PySTEMLab
nucleus marker, assisted by the automatic detector: its Bragg-contrast map and the objects into which it groups the
seed and grown pixels were displayed as a guide, and the outline of each nucleus was then drawn by hand; no automatic
object was taken over unedited (all saved masks are recorded as manual selections). These manual masks define the
nuclei and their 4-pixel ring in every analysis; the automatic objects are used, together with the manual ones, as
regions that random control placements avoid, so that controls contain no crystalline pixel.

\paragraph{Classifications.} $I(q)$ phase mapping (PySTEMLab phase mapping): the azimuthal average over integer
radial bins of $\log(1 + I)$ of the diffraction intensities, with the bright-field disc ($r < \SI{15}{px}$, its
edge after the Shift \& Mask centring; Supplementary Section~S5) masked,
PCA (30 components), three-dimensional UMAP\cite{mcinnes2018umap} (30 neighbours, minimum distance 0.1) and spectral clustering with an RBF affinity\cite{vonluxburg2007spectral}, six classes in
0033 and seven in 0034--0038. The number of classes was chosen by inspection of the class maps, because the silhouette
score is nearly flat over $k$ (it differs by at most 0.03 between six and seven classes in every scan; Supplementary
Section~S7). Whole-pattern PCA: every diffraction pattern of the window with the bright-field disc
($r < \SI{15}{px}$) and the corners beyond the last complete ring ($r > \SI{127}{px}$) masked, $\log(1 + I)$, PCA with
30 components, three-dimensional UMAP (30 neighbours, minimum distance 0.1) and spectral clustering with an RBF
affinity for six and seven classes; the number with the higher silhouette score was kept (seven in 0033 and 0038,
six in the other scans; silhouettes \numrange{0.49}{0.54}, within 0.02 of each other). These steps use the same
PySTEMLab routines as the phase-mapping dashboard. SimCLR: a foundational encoder trained by contrastive
self-supervised learning (SimCLR\cite{chen2020simclr}, NT-Xent loss with temperature 0.1) on 4D-STEM diffraction
patterns of chromium iron oxide, Ge--Sb--Te and MXene (\ce{Ti3C2T_x}) samples; a ResNet-style network (a $3 \times 3$ convolution with 64 filters and
eight residual blocks of 64--512 filters with stride-2 downsampling, global average pooling and a 128-dimensional
latent layer) with a two-layer projection head (64 dimensions), trained for 20 epochs with Adam (learning rate
\num{5e-4}, batches of 32) on views augmented by intensity scaling (0.8--1.2), Gaussian noise, random $90^\circ$
rotations and random $224 \times 224$ crops resized to $256 \times 256$. For inference in PySTEMLab the encoder was
converted to PyTorch (outputs identical to the original within \num{4e-7} relative); each pattern is divided by its
own maximum, the bright-field disc ($r < \SI{15}{px}$) is zeroed, and the 128-dimensional latent vector is the
embedding. The embeddings of the window were reduced with three-dimensional UMAP (30 neighbours, minimum distance
0.1) and clustered by spectral clustering with an RBF affinity for six and seven classes, keeping the number with the
higher silhouette score (seven in all scans except 0035; silhouettes \numrange{0.47}{0.54}); repeating UMAP and
spectral clustering with four other seeds reproduces the classes at ARI \numrange{0.69}{1.00}. Non-negative matrix factorisation (PySTEMLab component analysis) used three components. Agreement uses the adjusted Rand
index\cite{hubert1985ari} on probe positions assigned in both labelings; the depth-only partition divides the window
into six equal row bands. Enrichment compares the nucleus pixels (manual masks) of a class with the
class's share of the rows that contain nuclei (one-sided Fisher test).

\paragraph{Robustness of the Classifications.} Spatial coherence is the share of horizontally and vertically
adjacent probe positions with the same label, compared with the same labels shuffled within each row (20 shuffles),
which keeps the depth profile. Ring enrichment compares the share of the 4-pixel ring around the nuclei that falls in
a class with the expectation from the class's share of each row among the other window pixels. In the masked test all
crystalline pixels (manual and automatic objects), dilated by one pixel (63--150 probe positions per scan), are removed and the
classification is repeated with the same features, settings, seeds and number of classes; the interface class is the
class that holds most ring pixels, each nucleus object's site is the majority class of its own ring, and each removed
nucleus pixel is assigned the majority class of its 15 nearest kept pixels in the method's input space (the phase-mapping
features, the log pattern or the embedding). Refitting the $I(q)$ recipe on all pixels reproduces the published $I(q)$ classes
exactly in every scan (ARI 1.00).

\begin{figure}[!tbp]
  \centering
  \includegraphics[width=\linewidth]{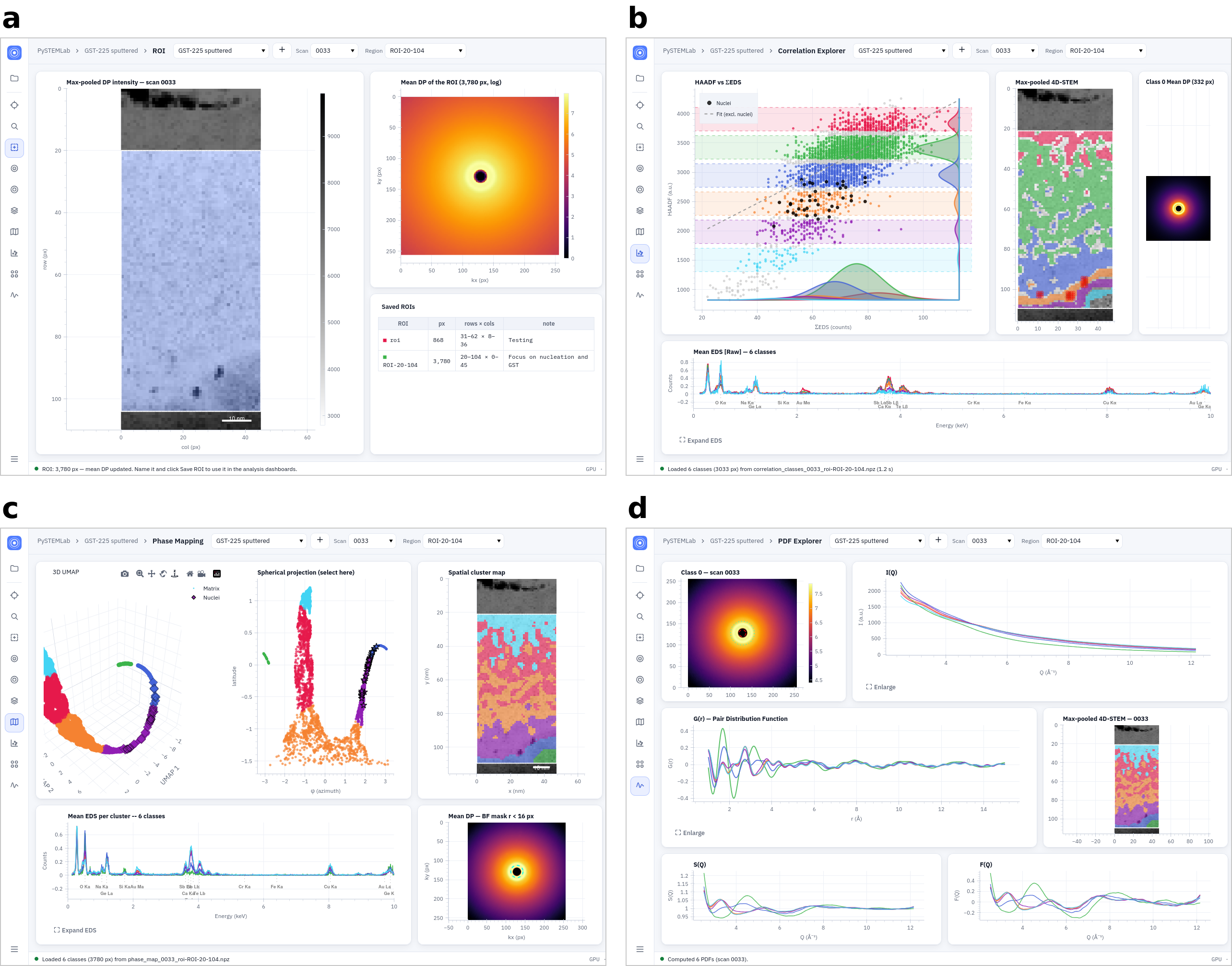}
  \caption{\textbf{PySTEMLab}, scan 0033, analysis window ROI-20-104 (working scan and region chosen once in the
  header for all dashboards). (a) ROI editor: the window drawn on the max-intensity virtual image, the mean diffraction
  pattern of the region and the saved ROIs. (b) Correlation explorer: HAADF versus \SEDS{} with the Theil--Sen fit
  (nuclei excluded) and the six HAADF bands, the bands in real space, the mean diffraction pattern of a class and
  the mean EDS spectra of all classes. (c) Phase mapping: three-dimensional UMAP of the $I(q)$ features, its spherical
  projection, the six $I(q)$ classes in real space, their mean EDS spectra and the mean diffraction pattern. (d) PDF
  explorer: $I(q)$, $G(r)$, $S(q)$ and $F(q)$ of the six $I(q)$ classes and their map. The navigation rail (left) and
  the shared project, scan and region selectors (top) are common to all dashboards; the settings dock on the right is
  not shown.}
  \label{fig:pystemlab}
\end{figure}

\paragraph{Pair Distribution Functions.} Mean patterns are computed exactly as weighted sums over all patterns and
integrated with pyFAI\cite{ashiotis2015pyfai} (256 points, polarisation 0.99) using PySTEMLab's PDF explorer, and
transformed with PDFgetX3\cite{juhas2013pdfgetx3} ($q_{\min} = 2.5$, $q_{\max} = q_{\max,\mathrm{inst}} =
\SI{12.17}{\qinv}$, $r_{\mathrm{poly}} = \SI{1.2}{\angstrom}$, composition from EDS: Ge 0.13, Sb 0.15, Te 0.32, O 0.32,
Si 0.045, Au 0.03). Reciprocal space is calibrated on the Au rings of the cap (Supplementary Section~S3). Peaks are the most prominent
local maxima in fixed windows (metal--oxygen 1.5--1.95, first shell \SIrange{2.45}{3.05}{\angstrom}). The FSDP is a
Gaussian on a linear background fitted in \SIrange{1.4}{2.8}{\qinv}, with the centre bounded to
\SIrange{1.75}{2.35}{\qinv}. Uncertainties are from 100 pixel-bootstrap replicates, each recomputing the mean pattern
exactly.

\paragraph{PySTEMLab.} All interactive analysis was carried out in PySTEMLab, a browser-based 4D-STEM analysis
environment written in Python (Panel, Bokeh) that we developed alongside this work (\cref{fig:pystemlab}). A project
file describes the scans, calibration and output folders; the data are read lazily from chunked zarr stores with
dask and, where available, a GPU, so that full $110 \times 45 \times 256 \times 256$ scans can be explored
interactively. Dashboards cover centring and masking (Shift \& Mask), inspection and virtual detectors,
region-of-interest selection shared across all dashboards, manual and automatic nucleus marking, HAADF--EDS
correlation with HAADF bands (correlation explorer), $I(q)$ phase mapping, embedding-based clustering,
non-negative matrix factorisation, and PDF extraction with calibration and bootstrap (PDF explorer). In this work the
analysis window was drawn in the ROI dashboard and every classification in the Results was made and exported from the
corresponding dashboard on that region; every interactive step is recorded in a session log, and the PDFs and
statistics were computed by scripts that call the same core routines with fixed parameters and seeds.

% ════════════════════════════════════════════════════════════════════════
\section{Results}

\subsection{Analysis Window and Dose}

The six scans (0033--0038) were recorded back to back on the same $110 \times 45$ pixel region in about one hour,
with \SIrange{88}{260}{\second} between scans (EMD metadata). At the recorded probe current of
\SI{0.14}{\pico\ampere} and \SI{100}{\milli\second} dwell, each scan delivers \SI{\approx 790}{e^- \per\angstrom\squared},
or \SI{4.7e3}{e^- \per\angstrom\squared} in total. The current was measured once, so the dose is an estimate.

All analyses use rows 20--103 of the scan, \SI{88}{\nano\metre} of the film (3780 probe positions per scan). Above
row 20 lies the Au cap, which we use only as an internal reciprocal-space calibration standard (Supplementary Section~S3); below row 104
lies the oxide layer, which is strongly beam damaged. The window is defined once as a region of interest and
used by every classification, and it contains every nucleus in every scan: the largest nucleus moves downwards
with dose and reaches row 103 in the last scan.

Over the series the window loses \SI{17}{\percent} of its mean HAADF intensity and \SI{13}{\percent} of its mean
\SEDS{}. Part of this drop may come from the probe rather than the film: the current was measured only once, and the
emission current of the cold field-emission gun decays continuously after the tip is flashed, so over the hour of the
series a decaying current would lower HAADF and \SEDS{} by the same factor everywhere. Such a decay cannot, however,
change the slope of the robust (Theil--Sen\cite{theilsen1950,sen1968slopes}) fit HAADF $= a\,\SEDS + b$, which falls
by \SI{27}{\percent}, or make the loss depend on depth, as it does, and the PDFs do not depend on the absolute
intensity. We therefore treat the series as a dose series of a beam-sensitive material\cite{bustillo2021beam}; an earlier clustering of the whole scan shows a damage front moving down from the Au
cap (Supplementary Section~S2). The beam also shortens the first shell and narrows the FSDP of the whole window
(Supplementary Section~S1 and Fig.~S1).

\subsection{Nuclei Lie on the HAADF--EDS Relation, and the EDS Spread is Counting Noise}

\begin{figure}[t]
  \centering
  \includegraphics[width=\linewidth]{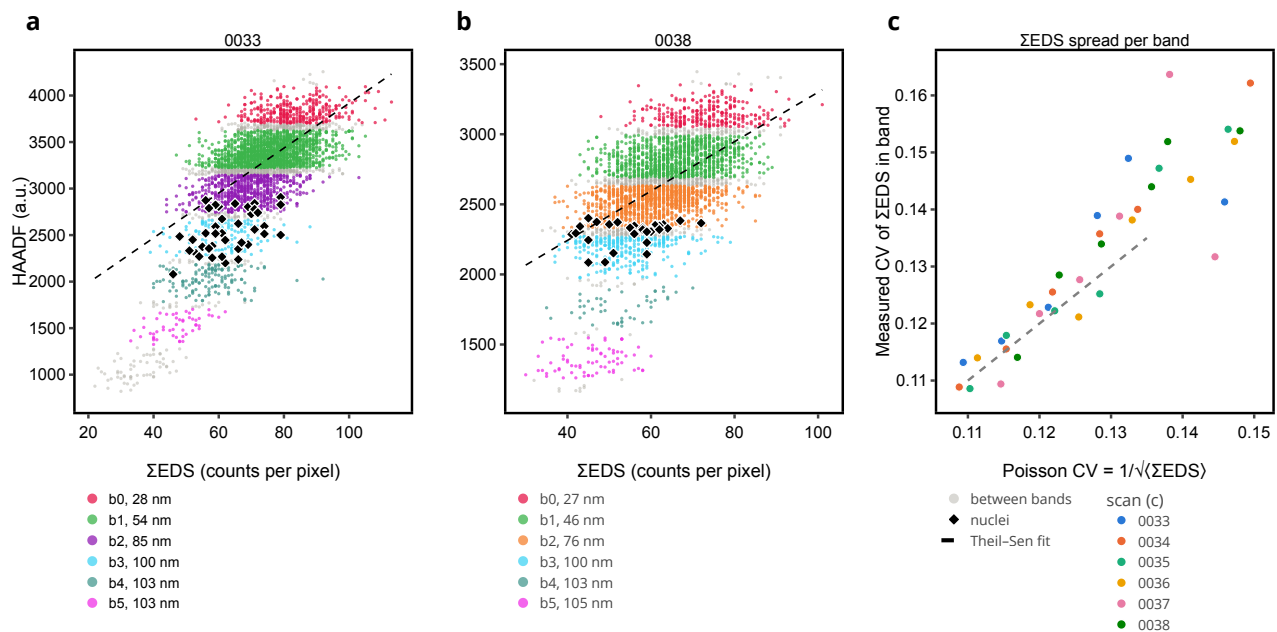}
  \caption{\textbf{HAADF versus summed EDS in the analysis window.} (a, b) Pixels of scans 0033 and 0038
  coloured by HAADF band (b0 brightest) in the depth colours of \cref{fig:classes} (key below each panel: band and
  mean depth), nuclei as black
  diamonds, Theil--Sen fit dashed. (c) Measured coefficient of variation of \SEDS{} within each band against the
  Poisson expectation, all six scans (colour: scan); the dashed line is equality.}
  \label{fig:haadfeds}
\end{figure}

If nucleation sites differed in density or composition, their pixels would lie off the HAADF--\SEDS{} relation of
the matrix. They do not. With the nuclei excluded from the Theil--Sen fit, the median residual of the nucleus pixels
is \numrange{-0.40}{-1.26} standard deviations of the residuals of matrix pixels at the same depth, within the
scatter of the matrix in every scan (\cref{fig:haadfeds}a,b, black diamonds). The residual is slightly negative in all six
scans and statistically detectable in five (Mann--Whitney $p = \num{6e-3}$ to \num{2e-9}; $p = 0.78$ in 0033), a
small HAADF deficit that we return to below, but the nuclei are not a separate population.

We then asked whether \SEDS{} even varies at fixed HAADF intensity. In PySTEMLab's correlation explorer we
divided the HAADF range into six bands of nearly constant HAADF intensity, i.e. of nearly constant mass--thickness
(the HAADF bands b0, brightest, to b5, darkest; \cref{fig:haadfeds}). Within every band of every scan, the coefficient of variation of \SEDS{} equals the
Poisson value $1/\sqrt{\langle\SEDS\rangle}$: the ratio is \num{1.03 \pm 0.05} over all 36 band--scan pairs. With only
\numrange{62}{72} X-ray counts per pixel, the \SEDS{} spread at fixed HAADF is counting noise, and no number-density
variation is resolved at the single-pixel scale. The bands therefore group probe positions by
mass--thickness; whether they also differ in composition or structure has to be read from the spectra and the
diffraction patterns (the PDFs below show a gradual change towards the oxide layer). A Gaussian mixture model on the HAADF--\SEDS{} plane, with the
number of components chosen by the Bayesian information criterion\cite{schwarz1978bic}, finds only two to four
broad, overlapping components that agree poorly with the bands (ARI \numrange{0.23}{0.34}): the HAADF--\SEDS{} plane
is a continuum without discrete populations. This is why we use fixed-width HAADF bands, a stratification for the
tests at fixed HAADF, rather than clusters chosen by a model-selection criterion: each mixture component spans a wide
HAADF range, within which the \SEDS{} spread would be dominated by the HAADF trend itself. Whatever distinguishes the
nucleation sites must therefore be read from the diffraction patterns.

\subsection{Nuclei are Crystalline and Visible Only in the Diffraction Signal}

Crystalline nuclei scatter into discrete Bragg spots on top of the amorphous rings. Because each pattern contains
\num{\sim 3e7} counts, Poisson noise is negligible and the coherent speckle of the amorphous matrix already exceeds
any Poisson threshold; we therefore measure the \emph{relative} contrast $c = (I - I_{\mathrm{ring}})/I_{\mathrm{ring}}$
of each local intensity maximum above the azimuthal mean at its radius. The strongest spot in amorphous patterns has
$c = \num{0.27}$ (robust s.d.\ \numrange{0.07}{0.08}), nucleus patterns reach $c \approx \numrange{0.8}{3}$, and
polycrystalline Au reaches $c \approx 3$. Pixels more than six robust standard deviations above each scan's own
background seed nucleus objects, which grow into neighbours above three standard deviations (hysteresis); no
threshold is set by hand.

\begin{figure}[p]
  \centering
  \includegraphics[width=0.49\linewidth]{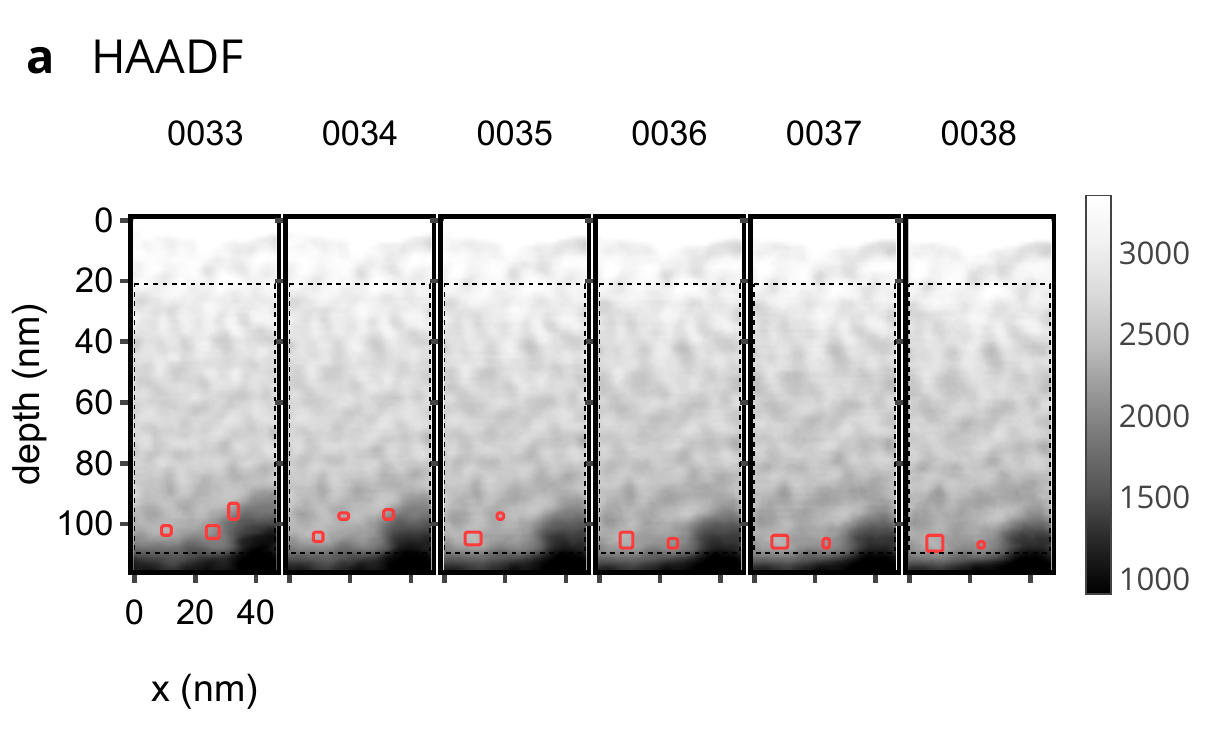}\hfill
  \includegraphics[width=0.49\linewidth]{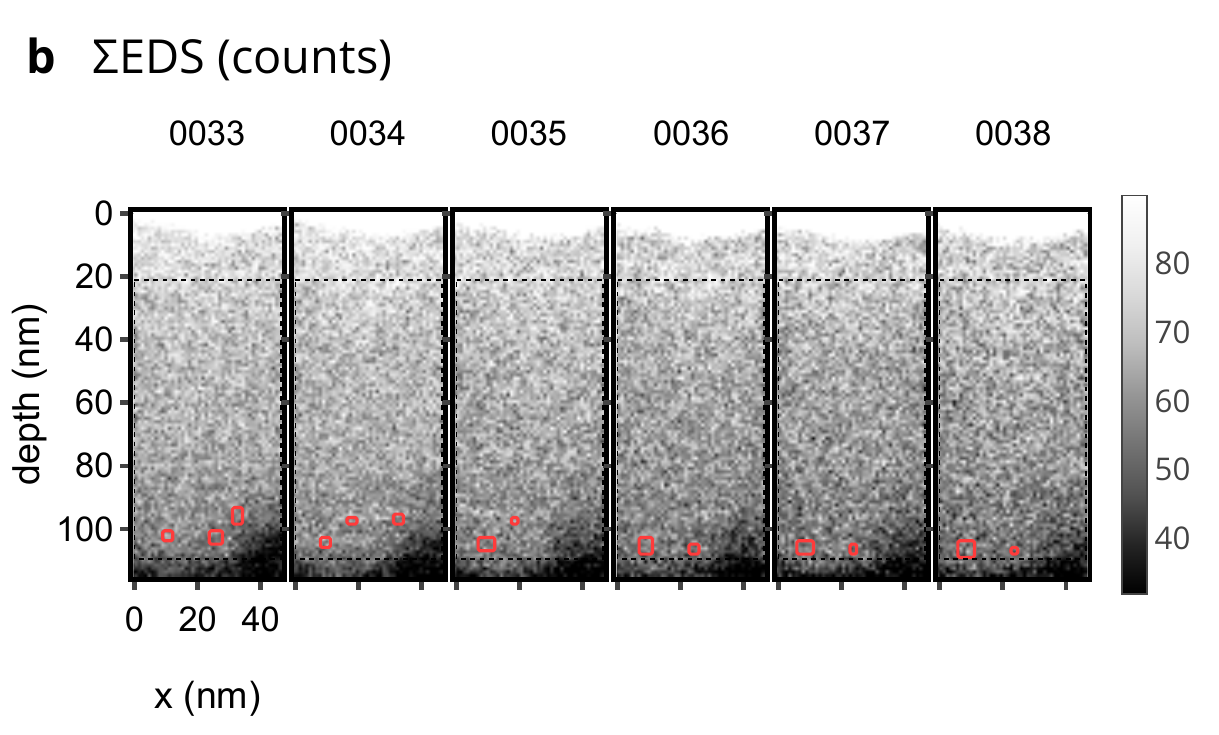}\\[2pt]
  \includegraphics[width=0.49\linewidth]{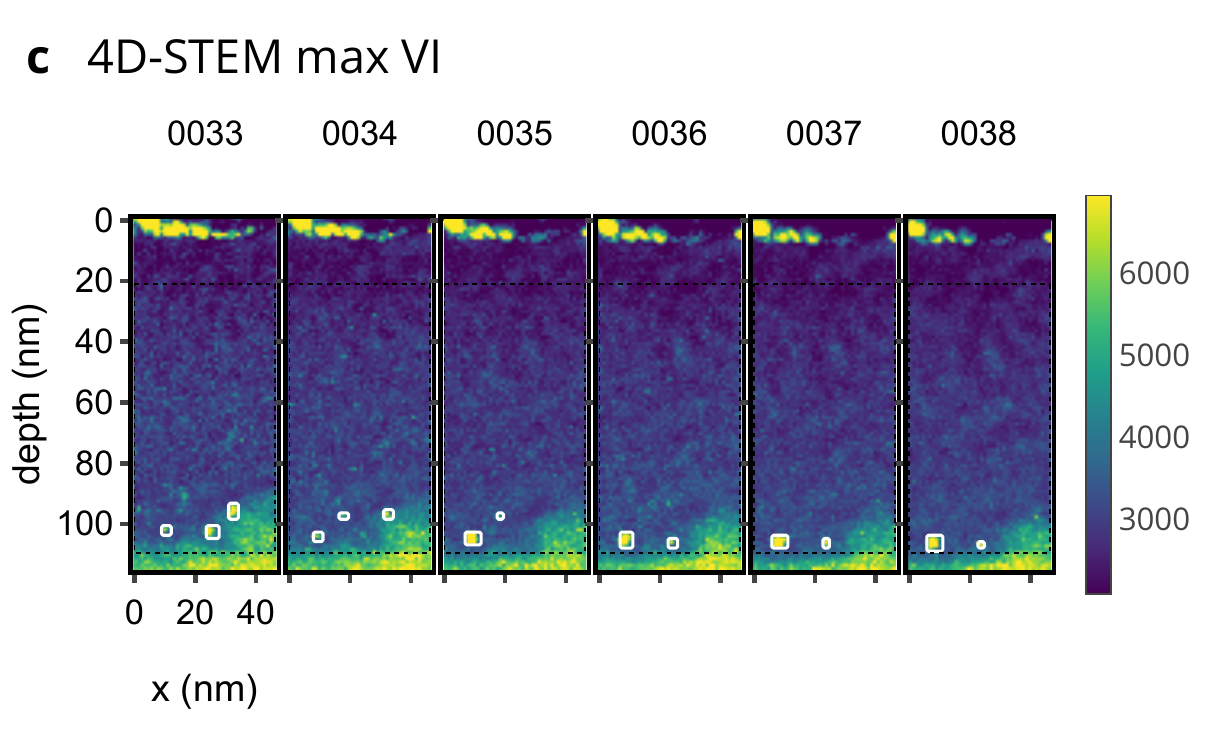}\hfill
  \includegraphics[width=0.49\linewidth]{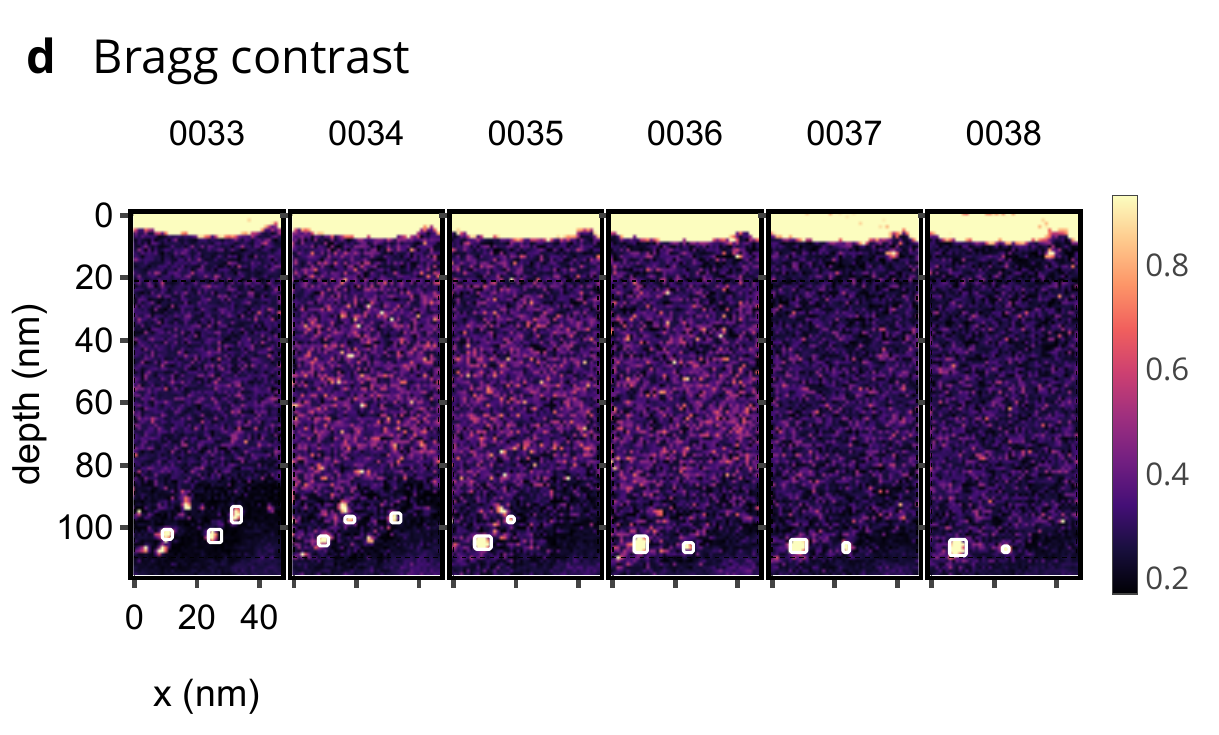}\\[4pt]
  \includegraphics[width=\linewidth]{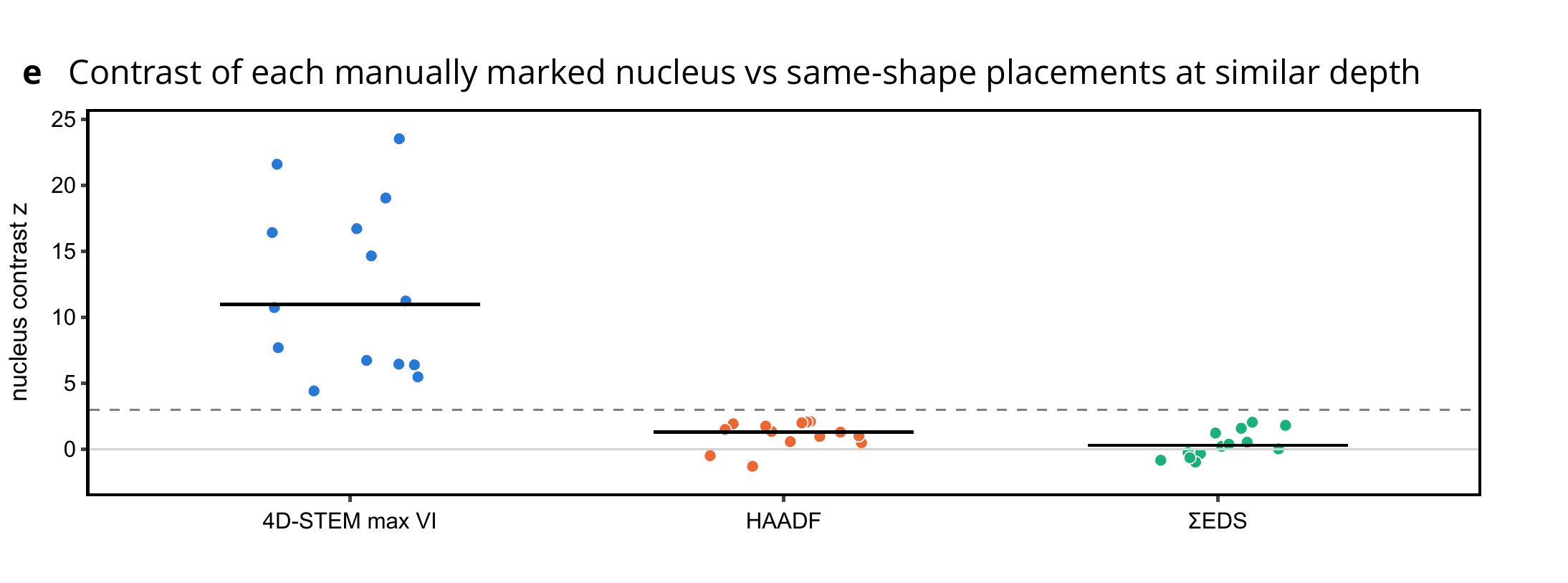}
  \caption{\textbf{Nuclei in four signals.} (a) HAADF, (b) \SEDS{}, (c) 4D-STEM max-intensity virtual image and
  (d) Bragg-spot contrast for each scan (full field of view; dotted line in a: top of the analysis window, which spans rows 20--103), with
  the manually marked nuclei outlined (red in a, b; white in c, d). (e) Contrast of each manually marked nucleus
  relative to its
  surroundings, as a $z$-score against all valid same-shape placements at similar depth inside the window; bars are
  medians, the dashed line is $z = 3$.}
  \label{fig:nuclei}
\end{figure}

The detector recovers all 14 manually marked nucleus objects across the six scans (\cref{tab:nuclei}) and finds
a few additional small objects. The manual masks (\cref{fig:nuclei}a--d) define the nuclei in all further
analyses. Because the manual marking was guided by the detector's Bragg-contrast map and objects (Methods), this
agreement is a consistency check rather than an independent validation; beyond it, the automatic objects are used
only to keep crystalline pixels out of the control placements. All manual nuclei and every automatic object larger
than three pixels lie at the GST--oxide interface (rows \numrange{86}{103}; five 2--3-pixel detections lie higher); the largest persists through all six scans while its peak Bragg contrast rises from \num{2.1} to
\num{3.0}, i.e.\ it becomes more strongly crystalline with dose. By contrast, eleven "nuclei" in the GST bulk from
an earlier intensity-threshold detection have spot contrast \numrange{0.28}{0.62}, within the amorphous background,
and none coincides with a detected object.

To ask which signals can see the nuclei, we compare each manually marked nucleus's contrast with its
surroundings against the same shape placed at every other valid position at similar depth inside the window. In
the 4D-STEM max-intensity virtual image all 14 objects exceed $z = 3$ (median $z = 11.0$, minimum $4.4$). In HAADF
none does (median $z = 1.3$, maximum $2.1$), and in \SEDS{} none does (median $z = 0.3$, maximum $2.1$)
(\cref{fig:nuclei}e). The nuclei are, to within these detection limits, crystalline
regions without a change in number density or composition that HAADF and \SEDS{} could resolve, consistent with
their position on the HAADF--\SEDS{} relation; the small HAADF deficit of their pixels (previous section) is
discussed in the Discussion.

\begin{table}[t]
\centering
\caption{\textbf{Nucleus detection per scan} (Bragg contrast; detection run on rows $\ge 14$, below the Au-rich
top of the cap; every object larger than three pixels lies in rows 86--103, inside the analysis window). Background: median and robust s.d.\
of the per-pattern maximum spot contrast. Objects: automatic and manual (max-intensity virtual image), with the
number of manual objects the automatic detector recovers and the pixel Jaccard index of the two masks.}
\label{tab:nuclei}
\small\setlength{\tabcolsep}{4pt}
\begin{tabular}{lcccccc}
\toprule
Scan & Background $c$ & Seed / grow & Auto obj.\ (px) & Manual obj.\ (px) & Recovered & Jaccard \\
\midrule
0033 & $0.276 \pm 0.084$ & 0.78 / 0.53 & 6 (60) & 3 (40) & 3/3 & 0.45 \\
0034 & $0.272 \pm 0.078$ & 0.74 / 0.50 & 7 (33) & 3 (24) & 3/3 & 0.46 \\
0035 & $0.274 \pm 0.076$ & 0.73 / 0.50 & 3 (30) & 2 (24) & 2/2 & 0.69 \\
0036 & $0.272 \pm 0.069$ & 0.69 / 0.48 & 5 (35) & 2 (29) & 2/2 & 0.64 \\
0037 & $0.275 \pm 0.070$ & 0.70 / 0.49 & 3 (41) & 2 (26) & 2/2 & 0.60 \\
0038 & $0.273 \pm 0.069$ & 0.69 / 0.48 & 4 (39) & 2 (29) & 2/2 & 0.62 \\
\bottomrule
\end{tabular}
\end{table}

\subsection{Four Classifications of the Film, and Where the Nuclei Fall}

\begin{figure}[p]
  \centering
  \includegraphics[width=0.49\linewidth]{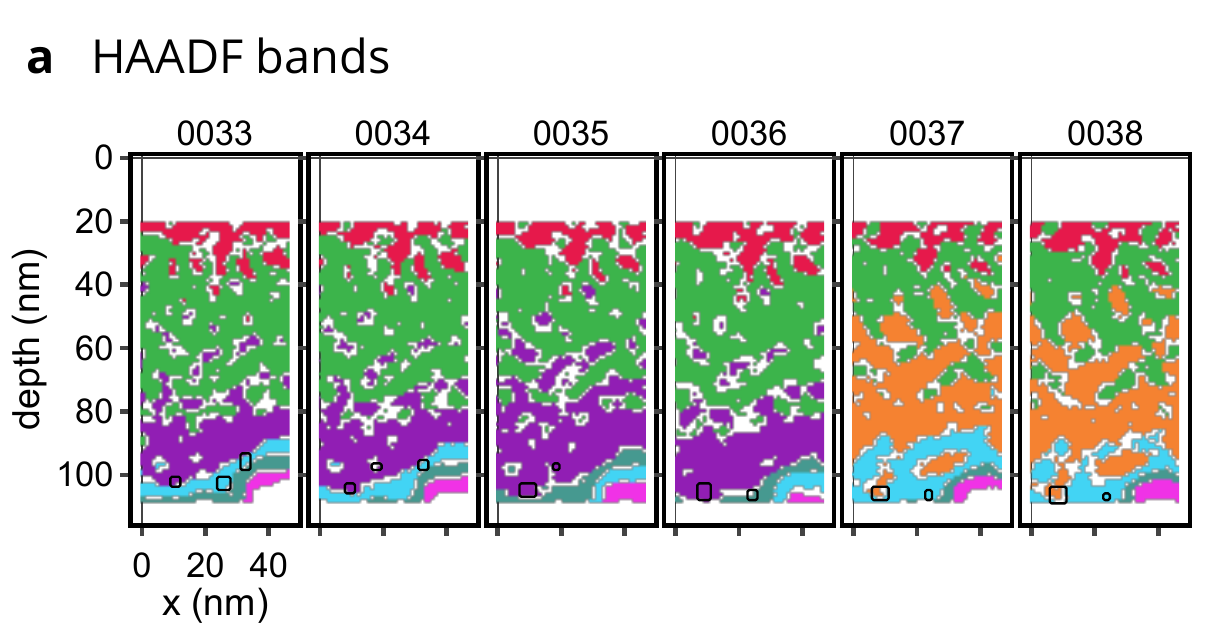}\hfill
  \includegraphics[width=0.49\linewidth]{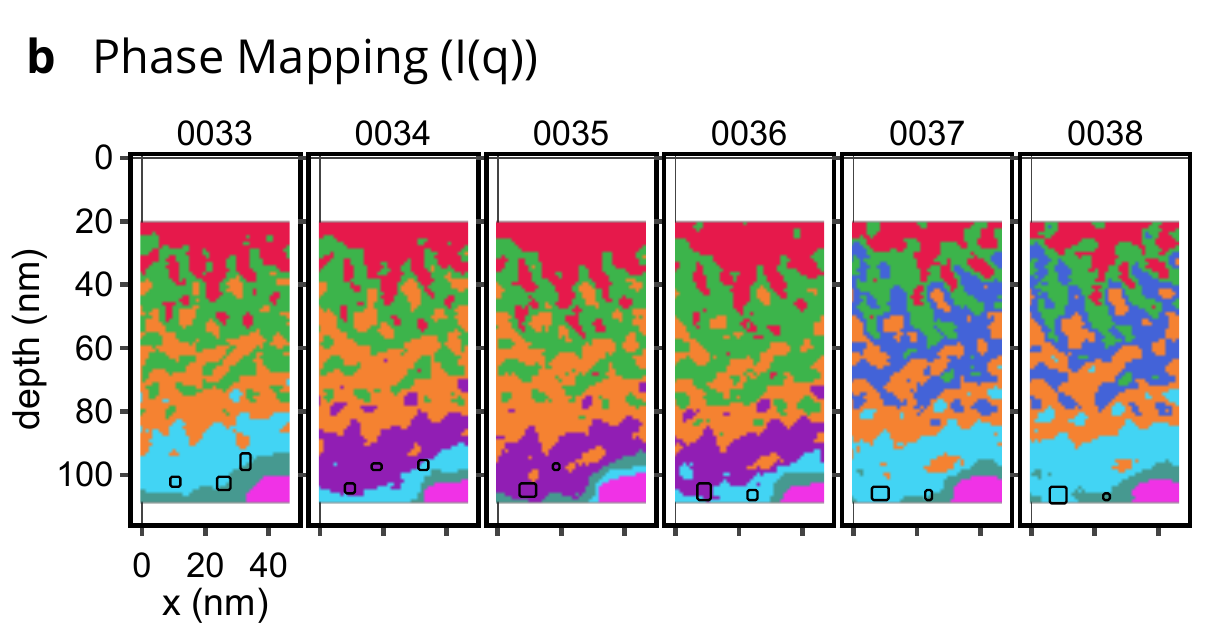}\\[2pt]
  \includegraphics[width=0.49\linewidth]{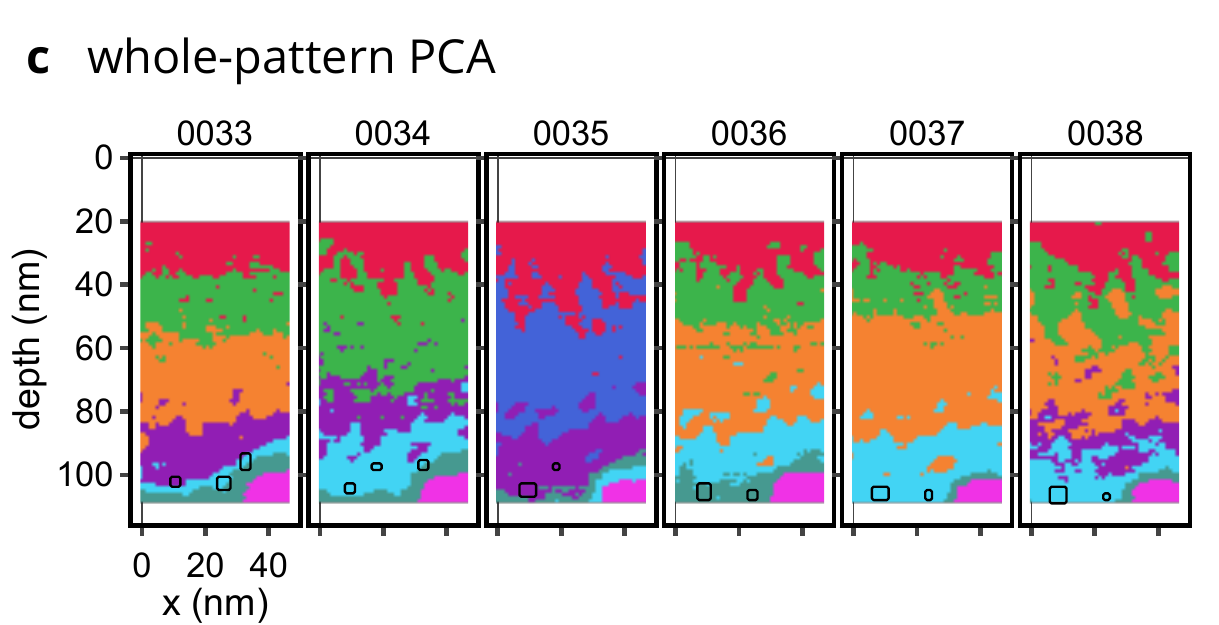}\hfill
  \includegraphics[width=0.49\linewidth]{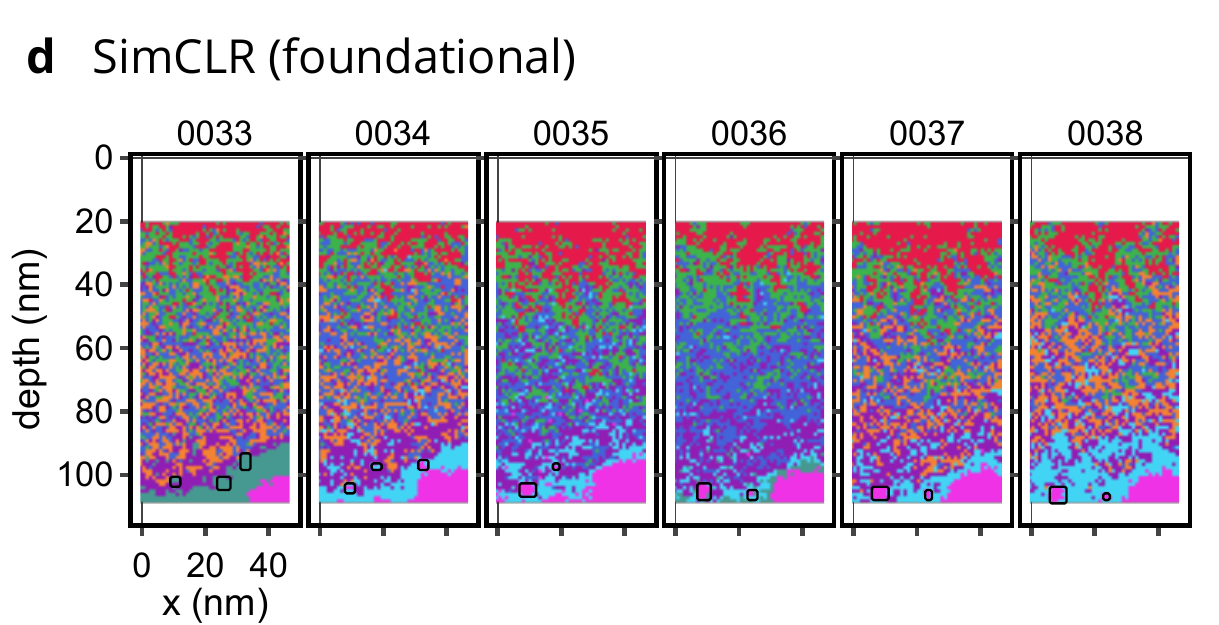}\\[4pt]
  \includegraphics[width=\linewidth]{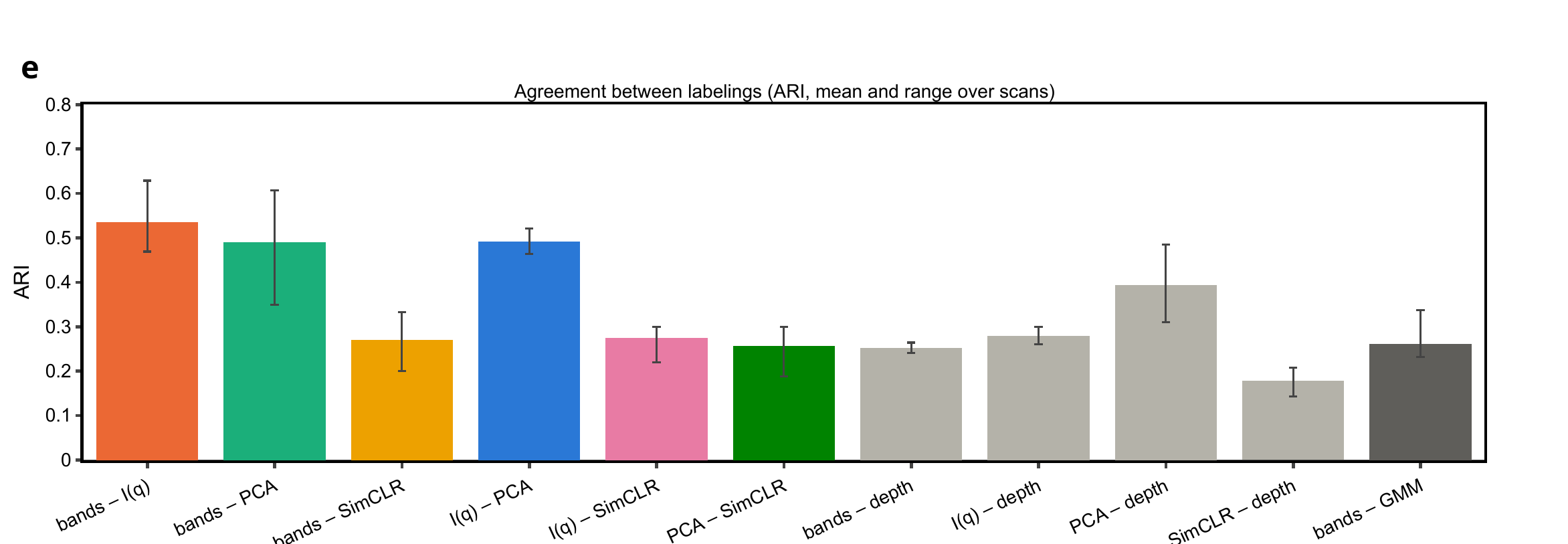}
  \caption{\textbf{Four classifications of the analysis window.} Class maps of (a) the HAADF bands, (b) $I(q)$
  phase mapping, (c) whole-pattern PCA and (d) SimCLR for every scan. Colours follow depth, the same in every method,
  scan and figure: each class takes the colour of the reference depth nearest its mean depth (29, 47, 62, 72, 83, 94,
  102 and \SI{106}{\nano\metre}: red, green, blue, orange, purple, cyan, magenta, teal), each colour used once per
  map; at the interface edge, where the mean depths coincide, magenta marks the class with the stronger metal--oxygen
  peak in $G(r)$ and teal the next. In 0033 the SimCLR classification has no near-interface bulk class (the
90--\SI{96}{\nano\metre} classes of the other scans, cyan), and its \SI{101}{\nano\metre} class already holds interface
material ($G = 0.17$ at the metal--oxygen peak, against 0.03--0.06 for those cyan classes), so it is teal. White: between bands;
nucleus outlines in black. (e) Agreement between the labelings (ARI, mean and
  range over the six scans), including a depth-only partition and a mixture model on HAADF--\SEDS{}.}
  \label{fig:classes}
\end{figure}

We classified the window in four different ways (\cref{fig:classes}a--d): by HAADF intensity (the six HAADF bands); by
phase mapping of the azimuthally averaged intensity $I(q)$ of every pattern (bright-field disc masked; PCA, UMAP and
spectral clustering into six classes in scan 0033 and seven in 0034--0038); by PCA of the whole diffraction patterns
followed by UMAP and spectral clustering; and by the 128-dimensional embeddings of a self-supervised (SimCLR)
foundational encoder (Materials and Methods), followed by UMAP and spectral clustering. In all three the
number of classes was six or seven, matching the six HAADF bands; for the last two the higher silhouette score
of the two decided per scan. The silhouette itself does not single out a number of classes: on the UMAP embeddings it
changes little between four and ten classes (\numrange{0.40}{0.60}) and peaks anywhere from three to nine classes
(Supplementary Section~S7). The classes therefore divide a continuous depth gradient rather than recover discrete
phases. A further decomposition, non-negative
matrix factorisation of the patterns, supports only two to three components and does not separate distinct regions.

The four labelings are related but not equivalent (\cref{fig:classes}e, \cref{tab:agreement}). The $I(q)$ classes
agree with the HAADF bands at ARI \numrange{0.47}{0.63}, well above the agreement of either with a depth-only
partition (\numrange{0.24}{0.30}): diffraction alone recovers the HAADF layering. The whole-pattern PCA classes
agree with both (\numrange{0.35}{0.61}) but follow depth most closely (\numrange{0.31}{0.48}), as expected when
depth-dependent changes of the total pattern dominate the principal components. The SimCLR classes agree only
moderately with the other three (\numrange{0.19}{0.33}) and least with depth (\numrange{0.14}{0.21}). This mostly
reflects their spatial noise: adjacent probe positions share a SimCLR class in only \SIrange{43}{52}{\percent} of
cases, against \SIrange{32}{42}{\percent} for the same labels shuffled within each row, whereas the HAADF-band,
$I(q)$ and PCA classes are spatially coherent (\SIrange{76}{91}{\percent} against \SIrange{47}{79}{\percent};
Supplementary Section~S7).

The classifications agree on where the nuclei are. In every scan the $I(q)$ classification has one class at the
GST--oxide interface (mean row 88--90) that holds \SIrange{62}{100}{\percent} of the nucleus pixels,
\numrange{1.2}{2.5} times more than its share of the nucleus rows alone (one-sided Fisher test, $p = \num{3e-2}$ to
\num{2e-9}). Whole-pattern PCA isolates such a class in every scan as well (\SIrange{55}{100}{\percent},
\numrange{1.4}{2.4}-fold, $p = \num{5e-3}$ to \num{1e-7}), the SimCLR classification in five of six and the
HAADF bands in four (\cref{tab:agreement}).

Because the nuclei are crystalline, a diffraction-based classifier could isolate them by their own Bragg spots. Two
tests show that the $I(q)$ and PCA interface classes are environments rather than nucleus classes (Supplementary
Section~S7). First, the 4-pixel ring of amorphous matrix around the nuclei is also enriched in them
(\numrange{1.12}{2.32}-fold for $I(q)$, \numrange{1.26}{2.21}-fold for PCA, depth-matched). Second, when every crystalline
pixel (manual and automatic objects) and a one-pixel margin are removed before classification, the classes hardly change (ARI
\numrange{0.93}{0.98} against the published ones), the interface class still forms (ring enrichment
\numrange{1.08}{2.71} for $I(q)$, \numrange{1.13}{2.50} for PCA), it surrounds 12 of the 14 nucleus objects in both,
and most removed nucleus pixels resemble it (\SIrange{38}{100}{\percent} and \SIrange{52}{100}{\percent} of them by
their 15 nearest neighbours). The SimCLR encoder behaves differently. Its
nucleus-holding class is small (\numrange{150}{355} probe positions at mean rows 91--100) and holds
\SIrange{58}{92}{\percent} of the nucleus pixels (\numrange{1.4}{2.2}-fold; $p = \num{3e-3}$ to \num{3e-10} in five
scans, $0.07$ in 0034) but, in four of six scans, not the ring around them (\numrange{0.00}{0.08}-fold in
0035--0038; \num{0.93} and \num{1.22} in 0034 and 0033): the encoder picks out the crystalline pixels themselves, a
label-free confirmation of the Bragg-contrast detection. With the nuclei removed, a SimCLR class surrounds the
nucleus sites (ring enrichment \numrange{1.19}{1.92}; 8 of 14 objects), but the removed nucleus pixels resemble it
much less (\SIrange{4}{42}{\percent}). The diffraction-based classifications thus isolate a nucleus-bearing interface
environment, which HAADF intensity alone does not, and the self-supervised encoder separates the crystalline nuclei from
that environment.

\begin{table}[t]
\centering
\caption{\textbf{Agreement between classifications} (ARI) and the nucleus-holding class of each method: share of
nucleus pixels (manual masks) / enrichment over the class's share of the nucleus rows (one-sided Fisher
$p$ in parentheses).}
\label{tab:agreement}
\scriptsize\setlength{\tabcolsep}{2.5pt}
\begin{tabular}{lcccccc}
\toprule
 & 0033 & 0034 & 0035 & 0036 & 0037 & 0038 \\
\midrule
bands -- $I(q)$                    & 0.54 & 0.47 & 0.52 & 0.63 & 0.54 & 0.53 \\
bands -- PCA                       & 0.43 & 0.61 & 0.35 & 0.47 & 0.54 & 0.54 \\
bands -- SimCLR                    & 0.20 & 0.28 & 0.26 & 0.33 & 0.26 & 0.29 \\
$I(q)$ -- PCA                      & 0.50 & 0.52 & 0.52 & 0.48 & 0.46 & 0.47 \\
$I(q)$ -- SimCLR                   & 0.22 & 0.27 & 0.28 & 0.30 & 0.28 & 0.29 \\
PCA -- SimCLR                      & 0.19 & 0.26 & 0.26 & 0.30 & 0.25 & 0.28 \\
bands -- depth only                & 0.25 & 0.26 & 0.25 & 0.26 & 0.25 & 0.24 \\
$I(q)$ -- depth only               & 0.30 & 0.29 & 0.28 & 0.28 & 0.26 & 0.26 \\
PCA -- depth only                  & 0.48 & 0.37 & 0.31 & 0.42 & 0.42 & 0.36 \\
SimCLR -- depth only               & 0.14 & 0.17 & 0.18 & 0.18 & 0.21 & 0.18 \\
bands -- GMM on HAADF--\SEDS{}     & 0.26 & 0.23 & 0.34 & 0.24 & 0.25 & 0.25 \\
\midrule
Band                               & \makecell{0.57 / 2.5\\(6e-7)} & \makecell{0.42 / 1.2\\(0.31)} & \makecell{1.00 / 2.8\\(2e-12)} & \makecell{0.69 / 1.8\\(4e-4)} & \makecell{0.35 / 0.7\\(0.96)} & \makecell{0.45 / 5.9\\(3e-10)} \\
$I(q)$ class                       & \makecell{0.80 / 1.2\\(0.03)} & \makecell{0.75 / 1.4\\(0.02)} & \makecell{1.00 / 1.7\\(1e-6)} & \makecell{0.62 / 2.5\\(5e-6)} & \makecell{0.96 / 1.8\\(3e-7)} & \makecell{0.86 / 2.4\\(2e-9)} \\
PCA class                          & \makecell{0.55 / 2.4\\(2e-6)} & \makecell{0.88 / 1.4\\(5e-3)} & \makecell{1.00 / 1.9\\(1e-7)} & \makecell{0.97 / 1.5\\(1e-5)} & \makecell{1.00 / 1.5\\(6e-6)} & \makecell{1.00 / 1.6\\(2e-7)} \\
SimCLR class                       & \makecell{0.85 / 2.2\\(3e-10)} & \makecell{0.58 / 1.4\\(0.07)} & \makecell{0.83 / 1.9\\(6e-5)} & \makecell{0.69 / 1.6\\(3e-3)} & \makecell{0.92 / 1.9\\(9e-7)} & \makecell{0.76 / 1.7\\(4e-4)} \\
\bottomrule
\end{tabular}
\end{table}

\subsection{Pair Distribution Functions of All Classes: One Short-Range Structure, a Depth Gradient and an Interface Edge}

\paragraph{Reliable Electron PDFs.} We calibrated reciprocal space against the six Au Bragg rings of the cap
(\SI{0.09511 \pm 0.00003}{\qinv} per pixel, \SI{34}{\percent} below the \SI{0.1451}{\qinv} of our earlier
studies; Supplementary Section~S3) and examined the processing. With the earlier settings every GST group showed a
dominant peak near \SI{3.7}{\angstrom}, which the crystalline Au cap does not show (Supplementary Section~S4). This
peak persists under changes of $q_{\max}$,
composition and polynomial order, and disappears once $q_{\min} \ge \SI{2.5}{\qinv}$: it arises from the
bright-field edge and from truncating the FSDP. With $q_{\min} = \SI{2.5}{\qinv}$, $q_{\max} = \SI{12.17}{\qinv}$
(the last complete detector ring) and the measured composition, the Au cap gives maxima at 2.78, 5.06 and
\SI{6.52}{\angstrom} (fcc Au: 2.88, 5.00 and \SI{6.45}{\angstrom}); its \SI{4.08}{\angstrom} shell is not resolved, and the
nearest-neighbour distance is \SI{0.1}{\angstrom} (\SI{3}{\percent}) short, so absolute distances carry a systematic
error of this order. Because the cut at $q_{\min}$ removes the FSDP from the Fourier transform,
medium-range order is measured directly in $I(q)$, by a Gaussian fit to the FSDP.

\begin{figure}[t]
  \centering
  \includegraphics[width=\linewidth]{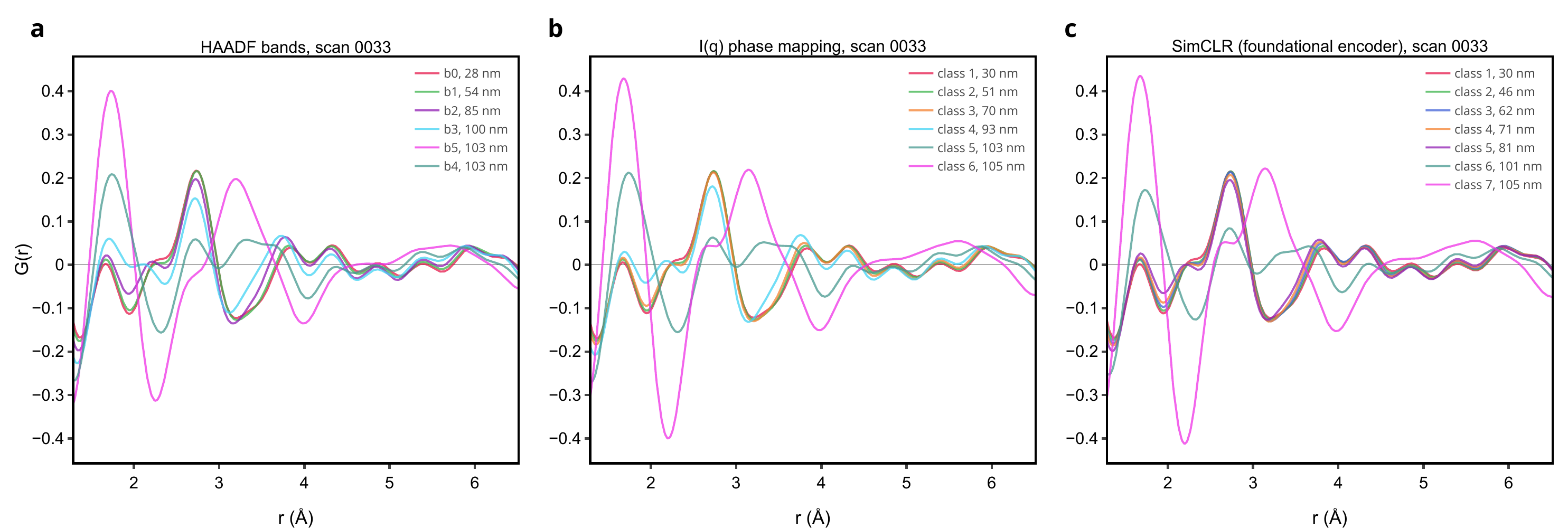}
  \caption{\textbf{Pair distribution functions of the classes of three classifications.} $G(r)$ of every class of (a)
  the HAADF bands, (b) the $I(q)$ phase map and (c) SimCLR in scan 0033, in the class colours of \cref{fig:classes}
  (key: class and mean depth). The curves were computed with PySTEMLab's PDF explorer pipeline from the mean diffraction pattern
  of each class over all its pixels, with the settings of \cref{tab:pdf} (Au-calibrated; $q_{\min} = \SI{2.5}{\qinv}$,
  $q_{\max} = q_{\max,\mathrm{inst}} = \SI{12.17}{\qinv}$, $r_{\mathrm{poly}} = \SI{1.2}{\angstrom}$, composition from
  EDS). The bulk classes overlap; the deviating curves are the
  interface-edge classes.}
  \label{fig:pdf}
\end{figure}

\paragraph{One Short-Range Structure.} We computed $G(r)$ and the FSDP of every class of all four methods in every
scan (25--27 classes per scan, 100 pixel-bootstrap replicates each; \cref{tab:pdf}; \cref{fig:pdf} shows the classes
of three of the methods in scan 0033). All classes
of more than 150 probe positions away from the bottom edge, whatever method defined them, have the same first shell,
\SIrange{2.70}{2.74}{\angstrom} (Ge/Sb--Te), with a metal--oxygen shell at \SIrange{1.61}{1.73}{\angstrom} consistent
with the \SI{32}{at.\percent} O measured by EDS. Within a scan the first shell of these bulk classes varies by at
most \SI{0.023}{\angstrom}.

\paragraph{A Depth Gradient in Medium-Range Order.} The FSDP position of the bulk classes falls with depth in every
scan, from \SIrange{2.059}{2.075}{\qinv} for the brightest band just below the Au cap (mean row 25) to
\SIrange{2.01}{2.05}{\qinv} for the deepest bulk classes near the interface (\cref{tab:pdf} lists the range over all
bulk classes). A lower FSDP position means a longer
characteristic medium-range distance; together with the longer metal--oxygen distance towards the interface in the
first scan (up to \SI{1.73}{\angstrom} in the deepest bulk classes against \SI{1.67}{\angstrom} near the top; in 0034 the
deepest classes are still up to \SI{0.01}{\angstrom} longer, and from 0035 on the difference reverses slightly, to
$\le$\SI{0.02}{\angstrom}), it is consistent with an increasing oxide contribution
towards the oxide layer, which is unambiguous only at the interface edge (below). HAADF bands resolve this gradient, because
they are stripes parallel to the layering, but no HAADF band above the interface edge has a distinct short-range
structure.

\paragraph{An Interface Edge Found by Every Method.} Each classification also returns a small class at the bottom
edge of the window (mean row $\ge 96$; together \numrange{631}{1241} probe positions per scan), the oxide side of the
interface. In every scan the deepest edge class of every method carries a metal--oxygen peak of $G = 0.20$--0.48,
against $\le 0.06$ in the bulk classes (0.17 in the SimCLR class of 0033 that reaches into the interface;
\cref{fig:pdf}), and its first shell is shorter (down to \SI{2.65}{\angstrom} in later scans) or not resolved at all.
The FSDP of the edge classes sits at low $q$, from \SI{2.04}{\qinv} down to the lower bound of the fit
(\SI{1.75}{\qinv}) in five of six scans. The nuclei straddle this edge and the interface class of the $I(q)$ phase map, i.e.\ they lie exactly where
the film changes from GST to the oxide-rich interface.

\begin{table}[t]
\centering
\caption{\textbf{Electron-PDF parameters per scan.} Window: the whole analysis window. Bulk classes: all classes of
the four methods with $> 150$ probe positions and mean row $< 96$ (15--20 per scan): ranges of the first shell $r_1$
and of the FSDP position. Edge classes: mean row $\ge 96$; $\le 1.75$: at the lower bound of the FSDP fit. Pixel-bootstrap s.d.: $r_1$
\SI{< 0.005}{\angstrom}, FSDP position \SI{< 0.005}{\qinv}, FSDP width of the window \SI{0.001}{\qinv}; these are
lower bounds, since the pixel bootstrap does not include the spatial heterogeneity of the film.}
\label{tab:pdf}
\small\setlength{\tabcolsep}{3.5pt}
\begin{tabular}{lcccccc}
\toprule
 & \multicolumn{3}{c}{Window} & \multicolumn{2}{c}{Bulk classes} & Edge classes \\
Scan & $r_1$ (\AA) & FSDP (\qinv) & FWHM (\qinv) & $r_1$ (\AA) & FSDP (\qinv) & FSDP (\qinv) \\
\midrule
0033 & 2.732 & 2.049 & 0.604 & 2.716--2.739 & 2.012--2.062 & $\le 1.75$--2.01 \\
0034 & 2.721 & 2.054 & 0.587 & 2.711--2.727 & 2.031--2.067 & $\le 1.75$--2.03 \\
0035 & 2.715 & 2.055 & 0.576 & 2.710--2.719 & 2.038--2.069 & $\le 1.75$--2.03 \\
0036 & 2.709 & 2.058 & 0.567 & 2.704--2.713 & 2.040--2.072 & 1.97--2.04 \\
0037 & 2.708 & 2.057 & 0.554 & 2.702--2.712 & 2.045--2.068 & $\le 1.75$--2.04 \\
0038 & 2.704 & 2.060 & 0.547 & 2.700--2.706 & 2.043--2.075 & $\le 1.75$--2.04 \\
\bottomrule
\end{tabular}
\end{table}

% ════════════════════════════════════════════════════════════════════════
\section{Discussion}

\paragraph{The Failed Hypothesis.} We started from the idea that nucleation sites differ in density or
composition. HAADF and \SEDS{} together are the direct test, and the test is negative: the nuclei lie within the
scatter of the matrix around the HAADF--\SEDS{} relation, they are invisible in \SEDS{} and barely visible in HAADF,
and the \SEDS{} spread at fixed HAADF is exactly the Poisson expectation. The small systematic HAADF deficit of
nucleus pixels relative to the fit may reflect diffraction contrast of the crystals at the high-angle detector
rather than a change in density, since \SEDS{} shows no matching deficit. Selecting probe positions by HAADF intensity
is likewise not enough: the bands follow the layering of the film but, down to the interface edge, are structurally
one material.

\paragraph{Nucleation at the Interface.} What does distinguish the nucleation sites is where they are. Every manually
marked nucleus and every automatically detected object larger than three pixels lies in the last
\SI{\approx 18}{\nano\metre} of the GST, at the transition to the oxide-rich
layer beneath it. The $I(q)$ phase map isolates this interface environment in every scan and places most
nucleus pixels in it, beyond what depth alone predicts, also when the nuclei are removed before classification; all four classifications separate the interface edge below it, with a
strong metal--oxygen peak, a shorter first shell and an FSDP at lower $q$; and the FSDP of the bulk classes moves to lower $q$ as the interface is
approached. The nucleating volume is thus defined by its environment, the GST--oxide interface,
and not by a composition or density that could be read from its own pixels.

\paragraph{Role of Oxygen.} The role of oxygen at such an interface is not fully understood, and the location of
the nuclei opens a route to explore it. Oxygen can drive phase separation: in refractory multiprincipal-element
alloys, oxidation creates chemical heterogeneities across length scales, from interstitial complexes to separate
oxygen-rich phases \cite{beaudry2024oxygen}. In non-oxide glasses, an oxide-rich phase or inclusion is a classic
heterogeneous nucleant. Oxide inclusions are the nucleants that must be dissolved or removed to approach homogeneous
nucleation in metallic glasses\cite{johnson2016quantifying}, and in a Zr-based metallic glass oxygen impurities switch
the dominant nucleation mechanism from homogeneous to heterogeneous, with an oxygen-rich phase crystallising
first\cite{tidefelt2024insitu}. The closest case to ours is Ge-rich Ge--Sb--Te. There, surface oxidation lowers the
onset of crystallization by \SIrange{50}{60}{\celsius} and moves it to the oxidised surface, from which it proceeds
inwards, and oxygen-bearing seeds such as \ce{Sb2O3} have been proposed as the heterogeneous nucleation
sites\cite{agati2020surface}. Our observations fit this picture: the nuclei form only at the transition to the
oxide-rich layer, the FSDP moves to lower $q$ towards it, and the interface-edge classes carry a metal--oxygen peak
several times stronger than the bulk. They do not identify the mechanism. At
the counting statistics of the present EDS data no compositional difference is resolved at the nuclei, so oxygen-bearing
seeds, an oxygen-driven phase separation of the amorphous film near the interface, or interfacial strain all remain
possible. Spatially resolved oxygen mapping across the interface, combined with the PDF classes used here, is the
natural next step to decide between them.

\paragraph{Beam-driven Ordering.} The dose series adds a second, unexpected result. While the beam removes material\cite{bustillo2021beam}
(HAADF $-17$\,\%, \SEDS{} $-13$\,\%, with a depth-dependent loss, of which a decaying probe current could explain
only a uniform part), the remaining film becomes more ordered: the first
shell shortens by \SI{0.03}{\angstrom}, the metal--oxygen shell by \SI{0.06}{\angstrom}, the FSDP narrows by
\SI{9}{\percent} and the correlation length grows from \num{10.4} to \SI{11.5}{\angstrom}, while the largest nucleus
grows in Bragg contrast. A preferential loss of oxygen or tellurium, which the EDS counts per pixel cannot resolve,
would shift the same peaks, so part of this change may be compositional rather than ordering. Quantitative nucleation statistics from 4D-STEM of GST at these doses must therefore be read
as dose-dependent, and low-dose acquisition is essential for nucleation studies\cite{bustillo2021beam} (Supplementary Section~S1).

\paragraph{Why the Pair Distribution Function.} The PDF is the measurement that connects the two results, the
environment in which nuclei form and the effect of the beam. HAADF reports
mean $Z$ and thickness, \SEDS{} the number of atoms and their kind, and a cluster label only that two groups of
patterns differ. The PDF gives the distances at which atoms sit and, through the FSDP, how far correlations extend,
on one quantitative scale for classes defined in completely different ways. It shows that HAADF-band, $I(q)$, PCA and
SimCLR classes share one short-range structure, that medium-range order changes with depth towards the interface,
that the interface edge differs in short-range order, and how the beam changes the film. Reliable electron PDFs
required two corrections, the Au calibration and the removal of the low-$q$ artefact, without which the dominant
peak of every class would have been a processing artefact at \SI{3.7}{\angstrom}.

\paragraph{The Role of Interactive Analysis.} Several results came from looking at the data in ways a fixed pipeline
would not have suggested: the HAADF bands were set in PySTEMLab's correlation explorer, the analysis window
was chosen once the Au cap and the damaged oxide layer had been identified on the maps, the classes of all methods were
inspected side by side on the same region, and the low-$q$ artefact was found by comparing the PDFs of the Au cap and
of GST. Each step was then reproduced by scripts with fixed parameters.

\paragraph{Limits.} The dose is estimated from a single current measurement; a decay of the cold-FEG emission
current during the series would make it an overestimate and would account for part of the uniform loss of HAADF and
\SEDS{}, though not for the change of their slope or the depth dependence of the loss. The analysis covers one region of one
sample, with 14 manually marked nucleus objects (two or three per scan), so per-scan nucleus statistics rest on few
objects. The electron PDF is processed
with X-ray-type normalisation and is used here for relative comparisons between groups processed identically; on
the Au cap it places the nearest-neighbour peak \SI{0.1}{\angstrom} short, so absolute distances, including the
first shell of \SIrange{2.70}{2.74}{\angstrom}, carry a systematic error of that order. This offset is not an error of
the $q$ scale, which the Au Bragg spots fix to \SI{0.04}{\percent} and which drifts by at most \SI{0.2}{\percent} between
scans (\SI{0.006}{\angstrom} at the first shell; Supplementary Section~S1), nor a single scale factor, since the longer
Au distances come out about \SI{1}{\percent} long. It is a bias of the processing, common to all groups processed
identically, and largely cancels in the differences between scans and classes on which the conclusions rest; the
dose-induced changes (\SI{1.0}{\percent} for the first shell, \SI{3.5}{\percent} for the metal--oxygen shell,
\SI{9}{\percent} for the FSDP width) are also not proportional, as a drift of the scale would make them. The bias
depends weakly on peak shape, which the beam also changes; this second-order effect is not quantified. The composition is taken as
constant across the dose series. The
SimCLR encoder was trained with each batch of 32 patterns divided by its maximum; applied in that way at inference it
smears the classes along the scan rows, so here each pattern is divided by its own maximum, an input distribution
the encoder was not trained on that may contribute to the spatial noise of its classes. Retraining with per-pattern
normalisation is planned.

\section{Conclusions}

\begin{itemize}
  \item Nucleation sites in this magnetron-sputtered GST film lie on the same HAADF--EDS relation as the matrix, and at
        fixed HAADF the summed EDS is at the Poisson limit: neither HAADF nor EDS distinguishes them.
  \item The nuclei are crystalline and visible only in the diffraction signal.
  \item HAADF-band, $I(q)$, PCA and SimCLR classes share one short-range structure; medium-range order changes with depth,
        and every method separates a distinct interface edge.
  \item The $I(q)$ phase map isolates the GST--oxide interface environment that holds most nucleus pixels in
        every scan, also when the nuclei are excluded from the classification, and a self-supervised encoder
        separates the crystalline nuclei from this environment.
  \item The electron beam removes material while shortening the first shell and narrowing the FSDP of the whole
        film over \SI{4.7e3}{e^- \per\angstrom\squared}.
\end{itemize}

\section*{Code and data availability}
PySTEMLab, the analysis scripts (with fixed seeds), all numerical results and the data are available from the
corresponding authors upon request.

\bibliography{references}

% Supplementary Information appended to the paper (same body as the standalone supplementary.tex)
\clearpage
\section*{Supplementary Information}
\setcounter{section}{0}\setcounter{figure}{0}\setcounter{table}{0}\setcounter{equation}{0}
\renewcommand{\thesection}{S\arabic{section}}
\renewcommand{\thefigure}{S\arabic{figure}}
\renewcommand{\thetable}{S\arabic{table}}
\renewcommand{\theHsection}{S\arabic{section}}   % unique hyperref anchors
\renewcommand{\theHfigure}{S\arabic{figure}}
\renewcommand{\theHtable}{S\arabic{table}}
\let\SIsection\section
\renewcommand{\section}{\FloatBarrier\SIsection}   % floats stay within their section, as placeins[section] in the SI
% Body of the Supplementary Information: \input by supplementary.tex (standalone SI)
% and by main.tex (SI appended to the paper for arXiv).

\section{Beam damage over the dose series}\label{sec:damage}

\begin{figure}[!htbp]
  \centering
  \includegraphics[width=0.86\linewidth]{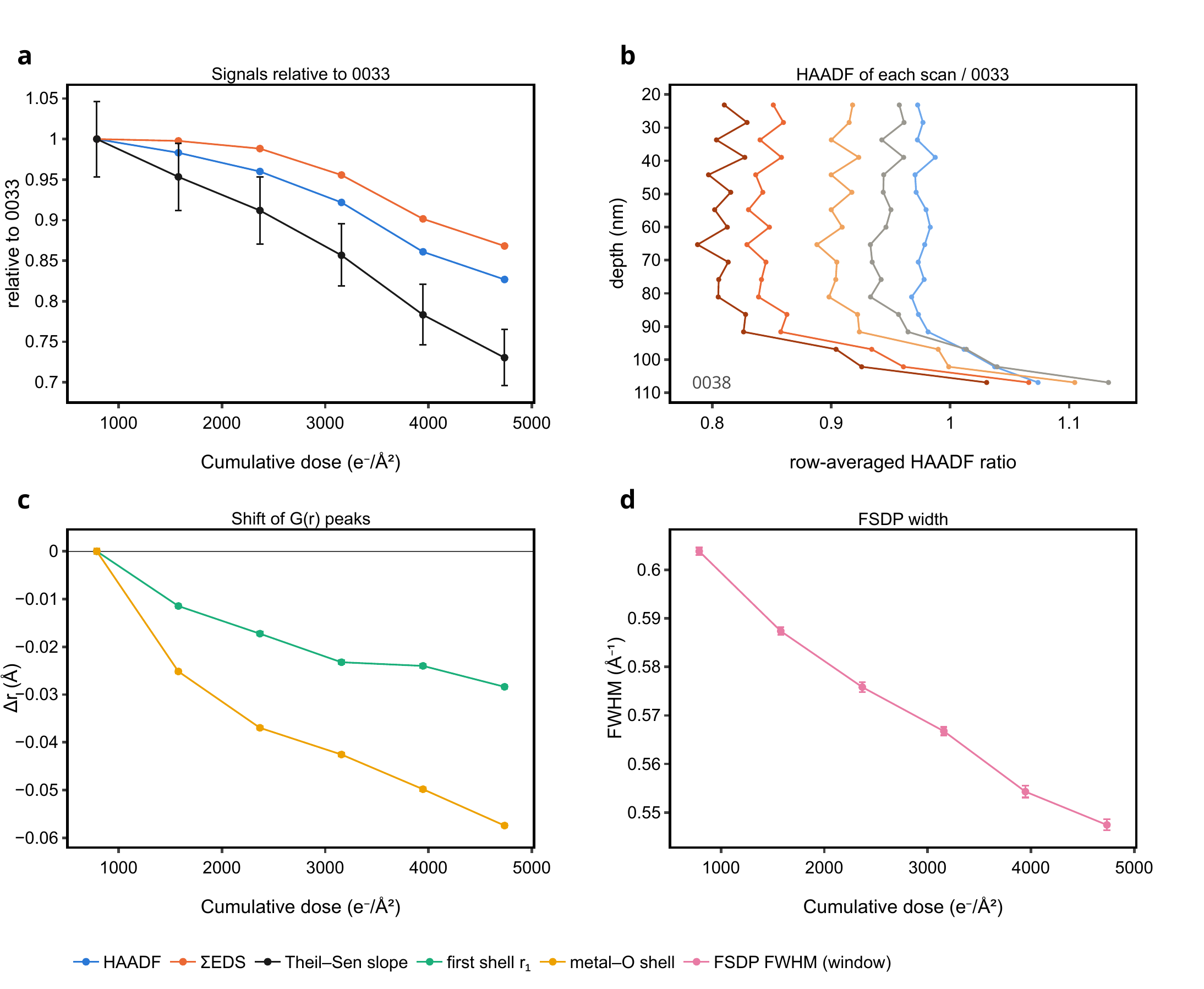}
  \caption{\textbf{Beam damage in the analysis window (rows 20--103).}
  (a) Mean HAADF and \SEDS{} and the Theil--Sen slope of HAADF versus \SEDS{} (nuclei excluded; bars:
  \SI{95}{\percent} CI), relative to scan 0033, against cumulative dose. (b) Row-averaged HAADF of each later scan
  divided by scan 0033, versus depth; a uniform current decay would give vertical lines. (c) Shift of the first
  shell $r_1$ and of the metal--oxygen shell of the window's $G(r)$ relative to 0033. (d) FSDP full width at half
  maximum of the window (bars: bootstrap s.d.).}
  \label{fig:damage}
\end{figure}

Over the series the window loses \SI{17}{\percent} of its mean HAADF intensity and \SI{13}{\percent} of its mean
\SEDS{}, and the slope of the robust (Theil--Sen) fit HAADF $= a\,\SEDS + b$ falls
by \SI{27}{\percent}, from \num{24.1} to \num{17.6} with non-overlapping \SI{95}{\percent} confidence intervals
(\cref{fig:damage}a). A decaying cold-FEG emission current, which cannot be excluded because the current was measured only once,
would scale HAADF and \SEDS{} by the same factor everywhere and leave $a$ unchanged; it may account for part of the
uniform drop, but not for the change of $a$ or for the depth dependence of the loss, with the last scan retaining between
\num{0.79} and \num{1.03} of the first scan's HAADF row by row (\cref{fig:damage}b). We therefore treat the series as
a dose series. The structural consequences, measured by the PDF of the whole window, are a first shell that
shortens from \num{2.732} to \SI{2.704}{\angstrom}, a metal--oxygen shell that shortens from \num{1.680} to
\SI{1.622}{\angstrom}, and an FSDP that narrows from \num{0.604} to \SI{0.547}{\qinv}, i.e.\ a correlation length
$2\pi/\mathrm{FWHM}$ that grows from \num{10.4} to \SI{11.5}{\angstrom} (\cref{fig:damage}c,d; bootstrap
uncertainties below \SI{0.001}{\angstrom} and \SI{0.001}{\qinv}). Fitting the Au calibration
(Section~S3) to each scan's Au-cap Bragg spots separately gives scales within \SI{0.2}{\percent} of each
other (0038 \SI{0.22}{\percent} below 0033), which would shorten the first shell by at most \SI{0.006}{\angstrom},
a fifth of the observed shift; a drift of the camera length therefore does not explain it. The beam thus removes material and, at the same
time, orders what remains (shorter bonds, narrower FSDP). The Au cap above the window changes as well: the first row with
row-mean Au below \SI{10}{at.\percent} moves from row 8 (0033) to rows 10--11 (0036--0038), and the median Bragg
contrast of the top rows falls from 3.3 to 1.8.

\section{Pooled clustering of all six scans}

This section comes from an earlier analysis of the whole scan (all 110 rows, including the Au cap and the oxide
layer), made before the analysis window was introduced. It supports the depth-dependent damage of
\cref{sec:damage}: a damage front moving down from the Au cap.
Clustering all $6 \times 4950$ patterns in one embedding separates the scans from one another when absolute
intensities are used. After normalising each $I(q)$ to unit sum, so that only its shape counts, three clusters
are very stable (ARI 0.97) and their fractions change smoothly with dose: an Au-cap-like shape spreads into the
upper GST (rows 12--40 from 31 to 94\,\%, rows 40--76 from 0 to 28\,\%), the oxide-like shape halves (14 to 7\,\% of
pixels), and the nuclei remain in the GST/interface cluster throughout.

\begin{figure}[!htbp]
  \centering\includegraphics[width=\linewidth]{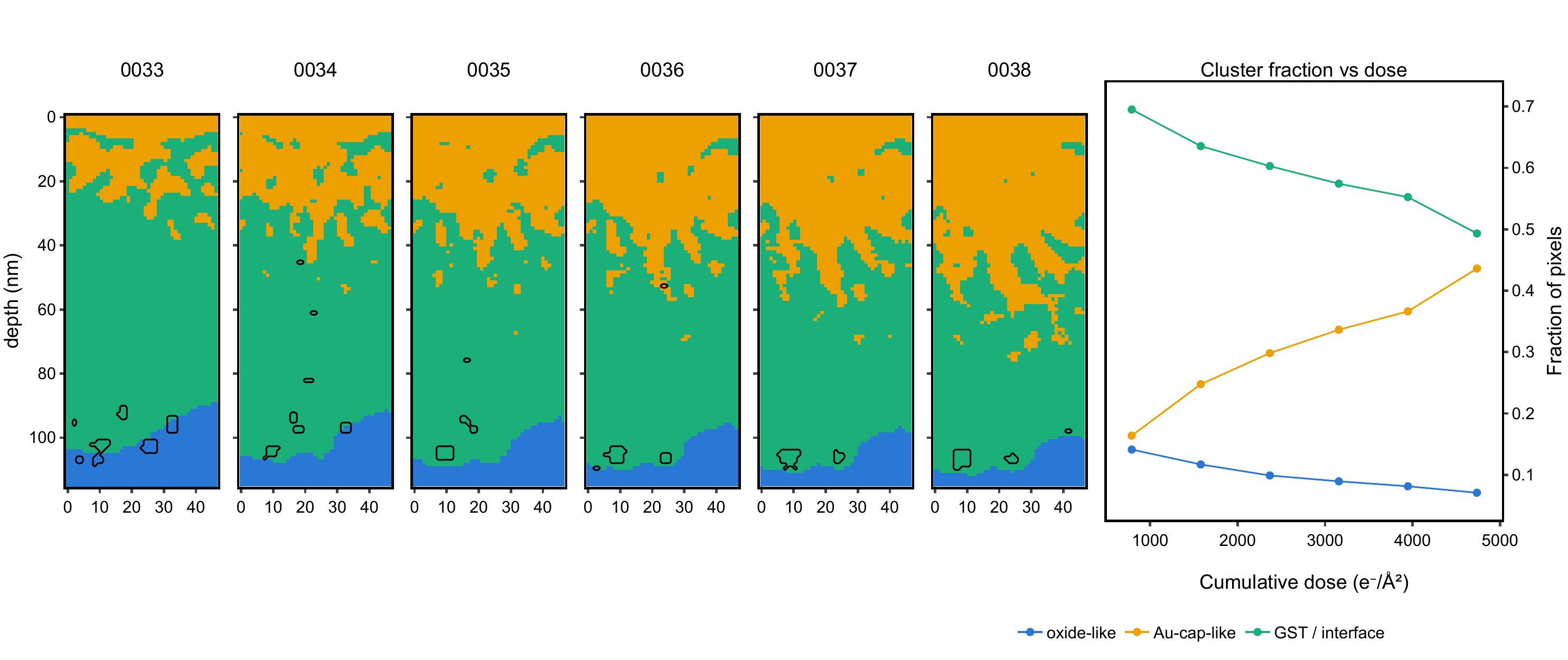}
  \caption{Shape-normalised pooled clusters for each scan and their pixel fractions against dose.}
\end{figure}

\section{Reciprocal-space calibration from the Au cap}

Bragg spots with relative contrast $c > 1$ in rows 0--5 of all six scans (7\,587 spots) were histogrammed by
radius. Six maxima index as fcc Au ($a = \SI{4.0782}{\angstrom}$) by radius ratio: (111), (200), (220), (311),
(400) and (420), at 28.08, 32.25, 45.82, 53.66, 65.16 and 72.23 px. A least-squares fit through the origin gives
\SI{0.09511 \pm 0.00003}{\qinv} per pixel (bootstrap over spots), with residuals of at most \SI{0.034}{\qinv}.
The PDF-explorer default initially used in this analysis (\SI{2.80}{\angstrom} at 22 px, \SI{0.1020}{\qinv}
per pixel) is \SI{7}{\percent} too large, and the \SI{0.1451}{\qinv} per pixel of our two earlier studies is
\SI{53}{\percent} too large; the equivalent PDF-explorer setting is $d_{hkl} = \SI{2.3594}{\angstrom}$ at a ring radius of 28 px. The
full detector ring ends at 128 px, i.e.\ $q = \SI{12.17}{\qinv}$.

\begin{figure}[!htbp]
  \centering\includegraphics[width=0.8\linewidth]{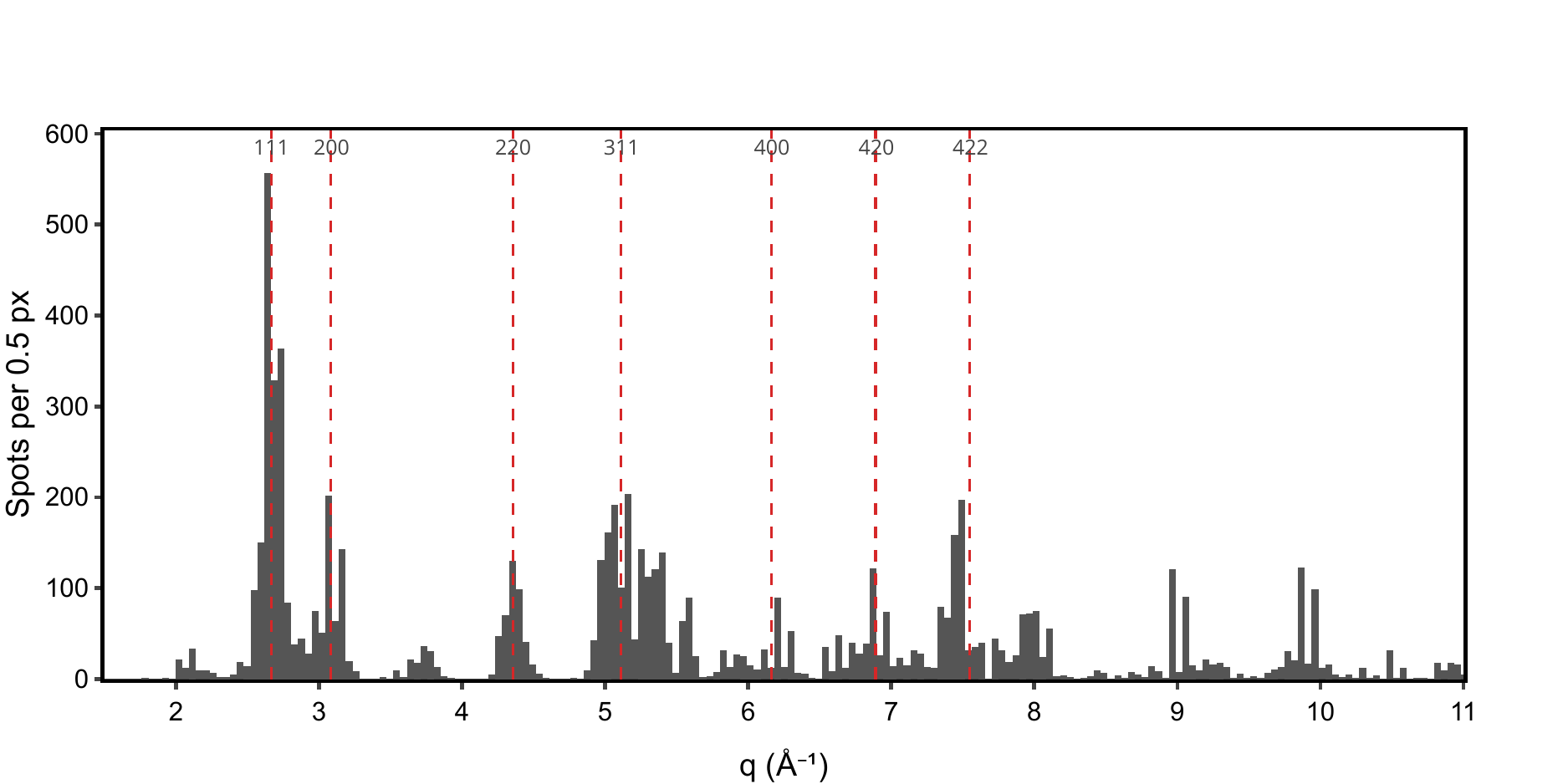}
  \caption{Radii of Au-cap Bragg spots converted to $q$ with the fitted calibration; dashed lines are fcc Au.}
\end{figure}

\section{Origin of the \texorpdfstring{$\sim$3.7\,\AA}{3.7 A} artefact in the electron PDF}

With the earlier settings ($q_{\min} = 1.17$, $q_{\max} = 15.2$, $q_{\max,\mathrm{inst}} = \SI{16.7}{\qinv}$,
$r_{\mathrm{poly}} = \SI{1.71}{\angstrom}$, composition \ce{Ge2Sb2Te5}) every GST group showed a dominant maximum at
\SIrange{3.65}{3.8}{\angstrom}. The crystalline Au cap, which has no diffraction peak near the FSDP (its first ring
is at \SI{2.67}{\qinv}), shows no such maximum: its $G(r)$ stays negative at \SIrange{3.3}{4.0}{\angstrom} for every
$q_{\min}$ tested (1.17--\SI{3.0}{\qinv}). In GST band 2 the ratio of the \SI{3.8}{\angstrom} feature to the first
shell is 1.9, 1.8 and 1.6 at $q_{\min} = 1.17$, 1.5 and \SI{2.0}{\qinv}, and 0.2 at \SI{2.5}{\qinv}. The
feature persisted when $q_{\max}$ (10--15.2), composition (nominal or measured) or $r_{\mathrm{poly}}$
(0.9--1.71) were varied, and collapsed only when $q_{\min}$ was raised to \SI{2.5}{\qinv} (in GST from
$G = 0.32$ to $0.045$). Two causes combine: the bright-field mask (8 px) is smaller than the contaminated edge of
the disc ($\approx$\SI{16}{px}, \SI{1.5}{\qinv}), and $q_{\min} = \SI{1.17}{\qinv}$ truncates the first sharp
diffraction peak (FSDP) near \SI{2.0}{\qinv}. The robust settings used in the main text ($q_{\min} = 2.5$,
$q_{\max} = q_{\max,\mathrm{inst}} = \SI{12.17}{\qinv}$, $r_{\mathrm{poly}} = \SI{1.2}{\angstrom}$, measured
composition Ge 0.13, Sb 0.15, Te 0.32, O 0.32, Si 0.045, Au 0.03) give Au maxima at 2.78, 5.06 and
\SI{6.52}{\angstrom} (fcc Au: 2.88, 5.00 and \SI{6.45}{\angstrom}); the \SI{4.08}{\angstrom} shell is not resolved,
and the nearest-neighbour peak is \SI{0.1}{\angstrom} short (\SI{2.79}{\angstrom} with the earlier settings).

\begin{figure}[!htbp]
  \centering\includegraphics[width=\linewidth]{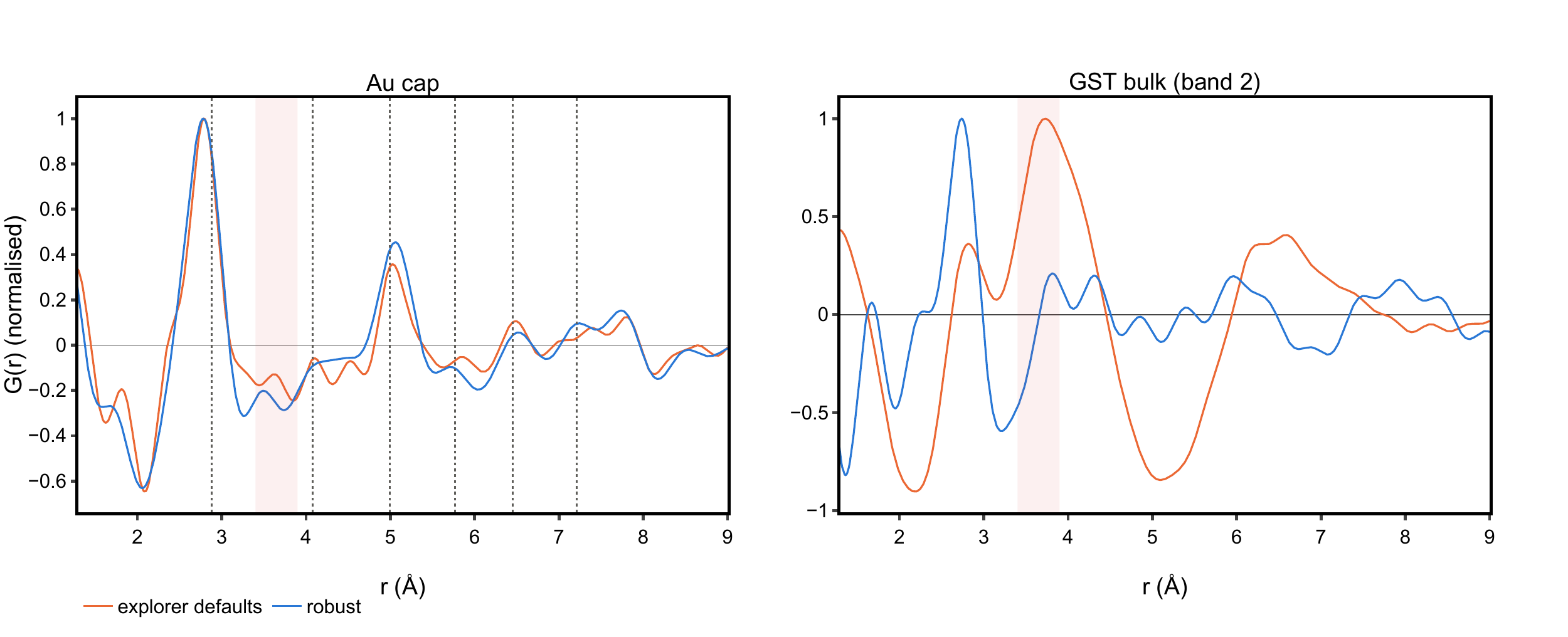}
  \caption{$G(r)$ of the Au cap (left; dotted lines: fcc Au distances) and of the GST bulk band (right) with the
  earlier PDF settings ($q_{\min} = \SI{1.17}{\qinv}$, $q_{\max} = 15.2$, $q_{\max,\mathrm{inst}} = \SI{16.7}{\qinv}$,
  $r_{\mathrm{poly}} = \SI{1.71}{\angstrom}$) and with the robust settings, both with the Au calibration. The shaded
  band marks the artefact region.}
\end{figure}

\section{Why radii below 16 px are excluded from clustering}

With all 128 radial bins, clusters in every scan reduce to the Au cap, the oxide layer, a fragmented "salt-and-pepper"
cluster (440--500 disconnected pieces) and one cluster holding 65--75\,\% of all pixels; agreement with the HAADF
bands is ARI $\approx 0.01$. Dropping bins below 16 px changes the labels almost completely (ARI between the two
versions 0.03--0.12), whereas replacing log by linear $I(q)$ does not (0.86--0.93). The low-radius bins sample the
edge of the masked bright-field disc and dominate the embedding. This earlier analysis excluded $r < \SI{16}{px}$;
the analysis of the rows 20--103 window masks $r < \SI{15}{px}$, the edge of the bright-field disc after the
Shift \& Mask centring, so both exclude the contaminated edge to within one pixel.

\section{Default versus hand-set HAADF band widths}

\begin{table}[h]
\centering
\caption{Agreement of the HAADF bands obtained with the band tool's default settings (six equal-width slots over the
12th--99.9th percentile span of HAADF, each band the central 78\,\% of its slot) with the bands set in the correlation
explorer with hand-set width and gap (pixels assigned in both; earlier analysis of the whole film below the cap).}
\begin{tabular}{lcccccc}
\toprule
Scan & 0033 & 0034 & 0035 & 0036 & 0037 & 0038 \\
\midrule
ARI & 0.76 & 0.94 & 1.00 & 1.00 & 0.39 & 1.00 \\
\bottomrule
\end{tabular}
\end{table}

\section{Robustness of the classifications}

\begin{figure}[!htbp]
  \centering\includegraphics[width=\linewidth]{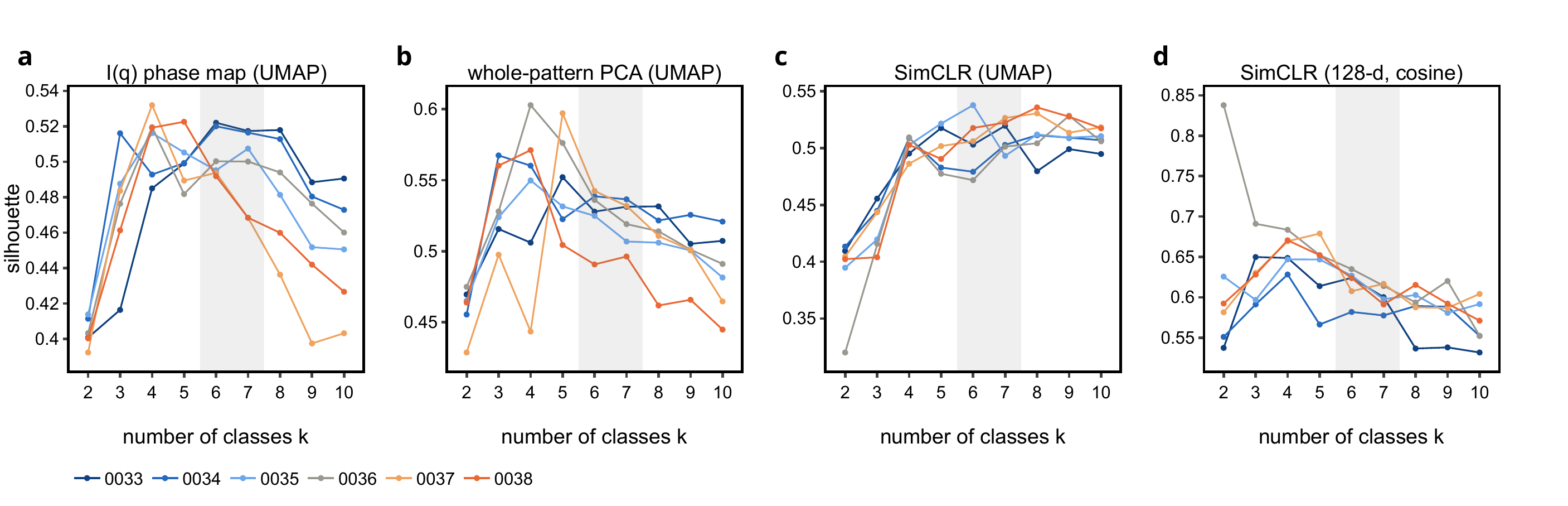}
  \caption{Silhouette score against the number of classes for each scan: (a) $I(q)$ phase map, (b) whole-pattern PCA
  and (c) SimCLR on their three-dimensional UMAP embeddings, and (d) SimCLR in its 128-dimensional embedding (cosine
  distance).
  Grey: the six or seven classes used.}
  \label{fig:kscan}
\end{figure}

\begin{figure}[!htbp]
  \centering\includegraphics[width=\linewidth]{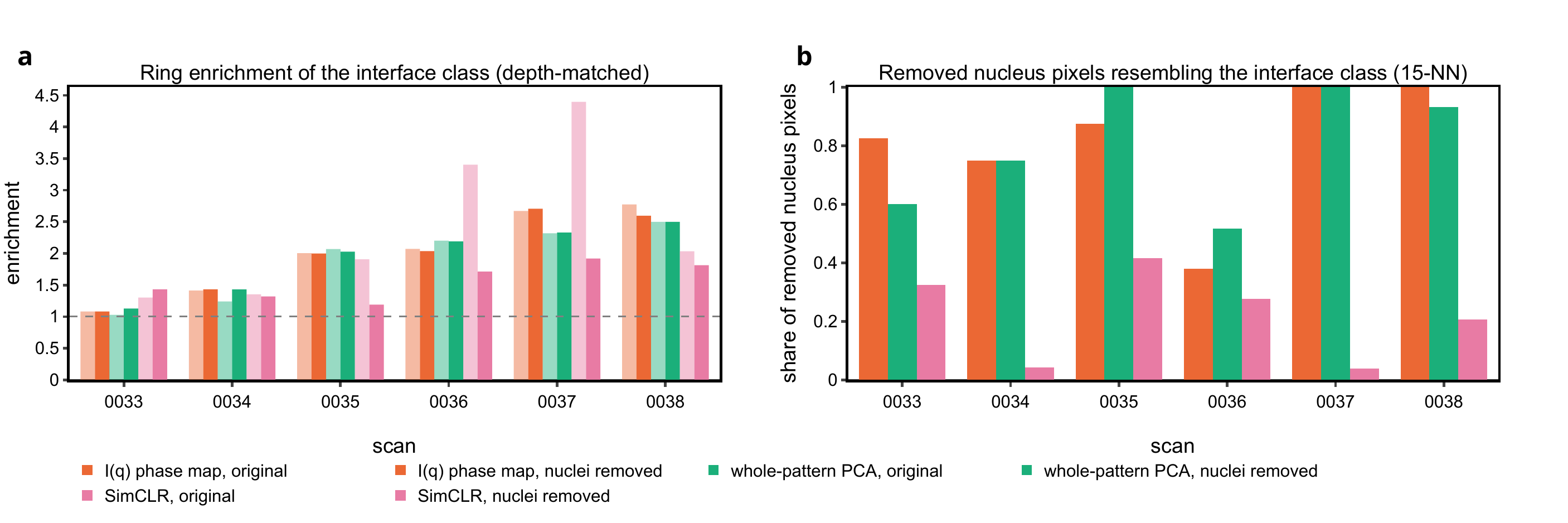}
  \caption{(a) Depth-matched enrichment of the 4-pixel ring around the nuclei in the interface class (the class
  holding most ring pixels), with the published classes (light) and with the nuclei removed (dark). (b) Share of
  the removed nucleus pixels whose 15 nearest neighbours lie mostly in that class.}
  \label{fig:maskedb}
\end{figure}

All numbers refer to the rows 20--103 window (\texttt{28\_cluster\_robustness.json}, \texttt{29\_masked\_nuclei.json});
definitions are given in the Methods of the main text. \Cref{tab:robust} summarises the range over the six scans.
The silhouette score changes little between four and ten classes and peaks at different numbers of classes in
different scans (\cref{fig:kscan}), so the number of classes (six or seven) is a choice, not a property of the data.
Removing the nuclei before classification leaves the $I(q)$ and PCA classes nearly unchanged and keeps an interface
class around the nucleus sites (\cref{fig:maskedb}); SimCLR also keeps a class around the sites, but the
removed nucleus pixels resemble it much less, because the encoder separates the crystalline pixels.

\begin{table}[!htbp]
\centering
\caption{Robustness of the four classifications (range over the six scans). Coherence: share of adjacent probe
positions with the same label, as published / shuffled within rows. Nucleus class: the class holding most nucleus
pixels (manual masks); depth-matched enrichment of the nucleus pixels and of the 4-pixel ring around them, with the
expectation taken row by row from the class's share of each row (larger than the enrichment of main-text Table~2,
which pools all rows that contain nuclei). Masked: classification
repeated without any crystalline pixel (manual and automatic objects, and a 1-pixel margin); ARI against the published classes, ring enrichment of the
interface class (ring 2--4 px from the nuclei, outside the removed margin), nucleus objects whose ring lies mostly in it, and removed nucleus pixels assigned to it by 15 nearest
neighbours.}
\label{tab:robust}
\scriptsize\setlength{\tabcolsep}{3pt}
\begin{tabular}{lccccccc}
\toprule
 & Coherence & \multicolumn{2}{c}{Nucleus class} & \multicolumn{4}{c}{Masked nuclei} \\
\cmidrule(lr){3-4}\cmidrule(lr){5-8}
Method & published / shuffled & nuclei & ring & ARI & ring & objects & removed px \\
\midrule
HAADF bands & 0.76--0.81 / 0.47--0.56 & 1.12--15.59 & 1.43--2.29 & -- & -- & -- & -- \\
$I(q)$ phase map & 0.82--0.85 / 0.49--0.58 & 1.49--5.84 & 1.12--2.32 & 0.93--0.98 & 1.08--2.71 & 12/14 & 0.38--1.00 \\
Whole-pattern PCA & 0.84--0.91 / 0.60--0.79 & 1.51--2.93 & 1.26--2.21 & 0.94--0.98 & 1.13--2.50 & 12/14 & 0.52--1.00 \\
SimCLR & 0.43--0.52 / 0.32--0.42 & 1.17--2.72 & 0.00--1.22 & 0.67--0.96 & 1.19--1.92 & 8/14 & 0.04--0.42 \\
\bottomrule
\end{tabular}
\end{table}

The HAADF bands are defined by HAADF, in which the nuclei are not visible, and were not re-classified. The
SimCLR classification is reproduced at ARI \numrange{0.69}{1.00} with four other UMAP and clustering seeds.

\FloatBarrier

\end{document}